\documentclass[12pt]{article}

\usepackage{scicite}

\usepackage{times}

\usepackage{color}
\usepackage{pgfplots}
\usepackage{epstopdf}
\usepackage{float}
\usepackage{titling}
\usepackage{hyperref}
\usepackage{paralist}
\usepackage{verbatim}

\newenvironment{sciabstract}{%
\begin{quote} \bf}
{\end{quote}}

\newcounter{lastnote}

\title{Fairy circle landscapes under the sea}

\author
{Daniel Ruiz-Reyn\'es$^1$, Dami\`a Gomila$^{1\ast}$, Tom\`as Sintes$^1$,
Emilio Hern\'andez-Garc\'{\i}a$^1$,\\ N\'uria Marb\`a$^2$, Carlos M. Duarte$^3$\\
\\
\normalsize{$^{1}$IFISC (Instituto de F\'isica Interdisciplinar y Sistemas Complejos)}\\
\normalsize{[Universidad Illes Baleares-Consejo Superior de Investigaciones Cient\'ificas (UIB-CSIC)].}\\
\normalsize{Campus Universitat Illes Balears, 07122, Palma de Mallorca, Spain}\\
\normalsize{$^{2}$Department of Global Change Research,}\\
\normalsize{IMEDEA (Mediterranean Institute for Advanced Studies) (UIB-CSIC),}\\
\normalsize{Miquel Marqu\'es 21, 07190 Esporles, Spain}\\
\normalsize{$^{3}$King Abdullah University of Science and Technology,}\\
\normalsize{Red Sea Research Center, Thuwal, 23955-6900, Saudi Arabia}\\
\\
\normalsize{$^\ast$To whom correspondence should be addressed; E-mail:  damia@ifisc.uib-csic.es}
}

\date{July 2017}

\begin{document}

% Double-space the manuscript.

\baselineskip24pt

% Make the title.

\maketitle

% Place your abstract within the special {sciabstract} environment.

\begin{sciabstract}
Short-scale interactions yield large-scale vegetation patterns that, in turn, shape ecosystem function across landscapes. Fairy circles, which are circular patches bare of vegetation within otherwise continuous landscapes, are characteristic features of semiarid grasslands. We report the occurrence of submarine fairy circle seascapes in seagrass meadows and propose a simple model that reproduces the diversity of seascapes observed in these ecosystems as emerging from plant interactions within the meadow. These seascapes include two extreme cases, a continuous meadow and a bare landscape, along with intermediate states that range from the occurrence of persistent but isolated fairy circles, or solitons, to seascapes with multiple fairy circles, banded vegetation, and \textit{"leopard skin"} patterns consisting of bare seascapes patterns consisting of bare seascapes dotted with plant patches. The model predicts that these intermediate seascapes extending across kilometers emerge as a consequence of local demographic imbalances along with facilitative and competitive interactions among the plants with a characteristic spatial scale of 20 to 30 m, consistent with known drivers of seagrass performance. The model, which can be extended to clonal growth plants in other landscapes showing fairy rings, reveals that the different seascapes observed hold diagnostic power as to the proximity of seagrass meadows to extinction points that can be used to identify ecosystems at risks.
\end{sciabstract}

% In setting up this template for *Science* papers, we've used both
% the \section* command and the \paragraph* command for topical
% divisions.  Which you use will of course depend on the type of paper
% you're writing.  Review Articles tend to have displayed headings, for
% which \section* is more appropriate; Research Articles, when they have
% formal topical divisions at all, tend to signal them with bold text
% that runs into the paragraph, for which \paragraph* is the right
% choice.  Either way, use the asterisk (*) modifier, as shown, to
% suppress numbering.

\section*{INTRODUCTION}

The spatial organization of vegetation landscapes is a key
factor in assessment of ecosystem health and functioning
\cite{Rietkerk2004,Barbier2006,Sole2006,Rietkerk2008}. Spatial
configurations of vegetation landscapes act as potential
indicators of climatic or human forcing affecting the ecosystem
\cite{Barbier2006} and determine energy and material budgets
and feedbacks across space
\cite{Barbier2006,Rietkerk2008,lefever1997origin,thiery1995model}.
In addition, although vegetation tends to cover all available
ground under favorable conditions, plant distributions can
spontaneously develop spatial inhomogeneities under resource
limitation, acting as ecosystem engineers that modify the
fluxes of nutrients and water to improve their growth
conditions \cite{Gilad2007,Rietkerk2008}. Hence, the spatial
organization of vegetation landscapes provides an indicator of
the existence of stressing factors and the proximity of
critical thresholds leading to irreversible losses
\cite{Rietkerk2008,Scheffer2009}.

The processes conducive to different dynamics of vegetation
landscapes have been assessed in habitats ranging from arid
ecosystems and savannahs to forests and wetlands
\cite{Rietkerk2004,Barbier2006,Rietkerk2008}. The most striking
patterns have been reported in drylands, where competition for
water leads to self-organized vegetation patchiness
\cite{von2001diversity,meron2004vegetation,Meron2012,fernandez2014strong,Scanlon07},
including the appearance of the so-called fairy circles, whose
origin has stirred significant controversy
\cite{cramer2013namibian,fernandez2014strong,juergens2013biological,Getzin16,tarnita2017}.
These are bare circular patches surrounded by grass, which
appear, for example, in regions of the grassy deserts of
Namibia \cite{cramer2013namibian} and Australia \cite{Getzin16}
and have been used as a basis to formulate a theory of
self-organization of vegetation landscapes in water-limited
ecosystems.

Although self-organized patchiness and pattern formation are
well documented in terrestrial ecosystems, their occurrence in
marine environments has attracted much less attention. Fairy
circle\^{a}��like structures have been reported in seagrass
meadows, such as Mediterranean \textit{Posidonia ocanica}
\cite{pasqualini1999environmental,bonacorsi2013posidonia} and
\textit{Zostera marina} in the Danish Kattegat
\cite{borum2014eelgrass}. Complex landscapes, such as bare
seascapes dotted with plant patches, termed \textit{"leopard
skin"} \cite{den1971dynamic}, and stripped vegetation patterns
\cite{frederiksen2004spatial,van2010spatial}, have also been
studied. However, inspection of satellite images and side-scan
cartography reveals that complex seascapes are abundant in
meadows of \textit{P. ocanica}, suggesting that self-organized
submarine vegetation patterns may be prevalent but have
remained thus far largely hidden under the sea. Obviously, the
mechanisms responsible for the formation of these submarine
fairy circle landscapes must be necessarily different than
those operating in water-limited ecosystems on land
\cite{cramer2013namibian,fernandez2014strong,Getzin16}.
Although there are some hypotheses of possible mechanisms in
several marine ecosystems
\cite{van2010spatial,marba1995coupling,marba1994migration,christianen2014habitat},
there is only a partial understanding of the phenomenon in
\textit{P. ocanica} meadows.

With a global distribution along the shorelines of all
continents except Antarctica, seagrass meadows are valuable
ecosystems that provide valuable ecosystem services
\cite{hemminga2000seagrass}; but they rank among the most
threatened ecosystems globally \cite{waycott2009accelerating}.
\textit{P. oceanica} is the dominant seagrass in the
Mediterranean Sea, where it forms underwater meadows that
support great biodiversity, are a site of intense CO$_2$
sequestration, and offer shoreline protection
\cite{duarte2013assessing,marba2014mediterranean}.
Unfortunately, this ecosystem is affected by multiple
anthropogenic impacts, including reduced water quality and
physical impacts, that have led to a loss of $6.9 \%$ per year
over the past 50 years \cite{marba2014mediterranean}. Although
\textit{P. oceanica} is a strongly clonal plant propagating
through rhizome growth, its very slow horizontal spread of a
few centimeters per year implies that losses are essentially
irreversible over managerial time scales
\cite{marba2014mediterranean}.

Here, we report that inspection of side-scan sonar cartography
of seagrass meadows (\textit{P. oceanica} and \textit{Cymodocea
nodosa}) in Mallorca Island (Western Mediterranean) reveals
that vegetation patterns of holes and spots spanning over many
kilometers are prevalent along the coastline of the Balearic
Islands (see Fig.~\ref{fig:PollensaAdriatic}a)
\cite{LIFE_Posidonia}. We also present a simple, parsimonious
model of clonal plant growth yielding self-organized submarine
vegetation patterns that encompass the diversity of seascapes
observed.

\section*{RESULTS}
Sintes et al. \cite{sintes2005,sintes2006modeling} developed a
model, based on a limited set of simple rules underpinning
clonal growth, successfully reproducing seagrass growth: First,
the growing apex of seagrass rhizomes elongates in a fixed
horizontal direction with a velocity $\nu$, leaving new shoots
behind, which are separated along the rhizome by a typical
distance $\rho$. Second, the growing apex develops new branches
at a rate $\omega_b$, with these branches elongating into a
horizontal direction at an angle $\phi_b$ from the original
rhizome. Living shoots have a typical lifetime, depending on
external factors and the presence of neighboring shoots,
resulting in a per capita mortality rate $\omega_d$. If
$\omega_d < \omega_b$ at a given position, the density of
shoots will increase locally. The typical value of each
parameter is a characteristic feature of each species with some
variability driven by genetic and environmental conditions
\cite{sintes2005}.

We have scaled up this model (originally conceived to describe
patch development) to landscape scale by coarse-graining it to
describe the dynamics of the shoot, $n_s$, and apex, $n_a$,
densities. Total shoot density $n_t$, is calculated as the sum
of shoots and apices growing in all directions. This model,
which we call the \^{a}�{\oe}Advection-Branching-Death\^{a}�� (ABD) model,
includes rhizome growth in different directions, contributions
from rhizome branching, and shoot death. Shoot mortality rates
are density-dependent as well as dependent on environmental
factors (for example, resource availability). More
specifically, three terms contribute to the total mortality
\begin{equation}
\label{EM3}
\omega_d[n_t(\vec{r},t)]= \omega_{d0}+ b n^2_t +\int \int
\mathcal{K}(\vec{r}-\vec{r}')(1-e^{-a n_t(\vec{r}')}) d\vec{r}'
\end{equation}

On the one hand, the intrinsic mortality rate, $\omega_{d0} >
0$, of an individual shoot at a particular position in the
landscape depends on environmental factors, and on the other
hand, on two density-dependent terms: local saturation and
nonlocal interaction.The saturation term $b n^2_t$ is nonlinear
and prevents an unlimited growth, increasing mortality locally
when the density increases excessively. The strength b of this
density-dependent term reflects the environmental carrying
capacity, determining the maximum density in the meadow.
Nonlocal interactions are included through an integral term
accounting for the interaction of shoots at position\^{a}��r with
those in a neighborhood weighted by the kernel
$\mathcal{K}(\vec{r}-\vec{r}')$. Through nonlocal interactions,
the abundance of shoots in a place can affect the growth in a
neighborhood. The model therefore includes two essential
components to yield self-organization: nonlinearity and spatial
interaction.

The three terms in Eq. \ref{EM3} are consistent with the
current functional understanding of seagrass meadows. The
intrinsic mortality rate of individual shoots, determining
$\omega_{d0}$, depends on external factors such as temperature
and irradiance regimes \cite{duarte1989temporal}. Local density
dependence results from self-shading, determining the maximum
density of shoots for a given plant size
\cite{duarte1987latitudinal} and the general decline in
seagrass density and biomass with depth
\cite{duarte1991seagrass}, as well as local depletion of other
resources, such as CO$_2$, which is depleted during daytime in
dense meadows \cite{invers1997effects}. Nonlocal interactions
integrate a number of facilitative and competitive mechanisms.
Facilitative interactions arise, for instance, in the form of
stress amelioration, when the presence of neighboring plants
dissipates wave energy, which may prevent shoot removal from
scouring of waves within the meadow
\cite{patriquin1975migration} and contributes to stabilize and
trap sediments \cite{gutierrez20117}. Facilitation has been
argued to play a role in shaping seagrass landscapes
\cite{fonseca2007biomechanical,bostrom2006seagrass}. Negative
interactions appear, for instance, as a result of anaerobic
microbial decomposition in dense meadows, which leads to
diffusing sulfide fronts that spread mortality, leading to the
appearance of fairy rings \cite{borum2014eelgrass}. Competition
can arise also from depletion of nutrients from the flow by
plants up-current \cite{cornelisen2006water} or of other
diffusing resources, such as CO$_2$.

Hence, existing evidence suggests intraspecific facilitation
and competitive nonlocal interaction whose precise ranges are
difficult to determine. As a result of these interactions, the
meadow can self-organize enhancing facilitative effects and
diminishing competition, altering the environment to yield more
favorable growth conditions.

We consider a kernel $\mathcal{K}$ with two terms of Gaussian shape
\begin{equation}
\label{EM4}
\mathcal{K}(\vec{r})=\kappa
\mathcal{G}(\sigma_\kappa,\vec{r})-\mu\mathcal{G}(\sigma_\mu,\vec{r})
\end{equation}

where $\kappa > 0$ is the strength of the competitive
interaction with width $\sigma_{\kappa}$, and $\mu >0$ is the
strength of facilitation with width $\sigma_\mu$, where the
widths of the Gaussians correspond to the spatial extension of
the interactions. Note that we should have $\mu \le
\omega_{d0}$ to guarantee positive mortality. For simplicity,
in the following, we take $\mu=\omega_{d0}$. As a result of the
two Gaussians with different widths and signs, the kernel has
the shape of an inverted Mexican hat, and the interaction is
stronger at short distances, decaying very fast with
$|\vec{r}|$.

Competition or facilitation may dominate at shorter or longer
distances depending on the values of $\sigma_\kappa$ and
$\sigma_\mu$. On general grounds, the main effect of
facilitation is to permit the coexistence of the populated and
unpopulated homogeneous states, whereas nonlocal competition is
one of the mechanisms responsible for the spontaneous formation
of regular patterns, either on itself \cite{MartinezGarcia2013}
or acting together with the facilitative interaction
\cite{Rietkerk2008}. Thus, observation of spatial patterns
suggests the existence of nonlocal competitive interactions.
Selecting $\sigma_\kappa > \sigma_\mu$ results in a kernel that
is weakly competitive at large distances, yielding to a
suitable nonlocal interaction for pattern formation (see the
Supplementary Materials) \cite{Rietkerk2008}. The main nonlocal
interaction terms ($\kappa$, $\sigma_\kappa$, $\mu$,
$\sigma_\mu$) can be inferred from the comparison of numerical
simulations and observed patterns, whereas seagrass growth
parameters are largely known [see the study by Sintes et al.
\cite{sintes2006modeling} and references therein].

The ABD model yields a great diversity of complex spatial
patterns emerging at different parameter regions, including
nonlinear phenomena resulting in the coexistence of different
solutions, which are shown in Fig.~\ref{fig:biffurcation} for
the case of constant $\omega_{d0}$ and $b$. When mortality is
high ($\omega_{d0}/\omega_b >>1$), the only possible solution
is bare soil, the unpopulated solution. Decreasing the
mortality (or increasing branching rate), the unpopulated
solutions become unstable at a threshold
$\omega_{d0}/\omega_b=1$. Below this mortality, any small
nonzero density will grow to form a meadow. If mortality is
much smaller than the branching rate, $\omega_{d0}/\omega_b <<
1$, the vegetation will uniformly cover all the available
space. Thus, there are two extreme homogeneous stable states: a
uniform, continuous meadow ($\omega_{d0}/\omega_b << 1$) and
bare seafloor, with no vegetation ($\omega_{d0}/\omega_b >>1$),
as it is shown in Fig.~\ref{fig:biffurcation} (red solid
lines). The parameter values that better reproduce the observed
patterns of \textit{P. oceanica} meadows correspond to a kernel
$\mathcal{K}$ that, although competition extends farther, is
overall facilitative. As a result, the transition (T) from bare
soil to the populated solution is subcritical, so that there is
a mortality range in which both populated and unpopulated
solutions coexist (Fig.~\ref{fig:biffurcation}). Competition
between shoots can destabilize the populated solution, leading
to patterns, whereas the homogeneous states are the only
possible solutions in the absence of nonlocal interactions.
Therefore, the presence of fairy circle landscapes in seagrass
meadows is indirect evidence of nonlocal negative interactions.

A linear stability analysis of the homogeneous populated
solution reveals that it undergoes a finite wavelength
instability, also known as Turing \cite{Rietkerk2008} or
modulation instability (MI), at a critical value of mortality
rate $\omega_{d0}=\omega_{d0}^c$ (in Fig.
\ref{fig:biffurcation}, $\omega_{d0}^c/\omega_b =1.34$),
leading to the emergence of complex spatial patterns. Above
this mortality, any small perturbation to the homogeneous
meadow is enough to trigger a feedback process driving the
vegetation to form an inhomogeneous pattern. Different spatial
structures are possible. As mortality rate increases, possible
patterns shift from negative hexagons (holes arranged in a
hexagonal pattern), to stripes, and to positive hexagons (spots
of vegetation arranged in a hexagonal pattern), as expected
from the general theory of pattern formation \cite{Walgraef,
Silber15}. Each solution is stable in a different region of
parameter space, and two different solutions can be
simultaneously stable for the same value of the mortality, a
coexistence between solutions that would give rise to
hysteresis. In particular, negative hexagons coexist with the
homogeneous populated solution in a mortality range below
$\omega_{d0}^c$. In part of this region, a single bare hole
embedded in a dense meadow (Fig. ~\ref{fig:hole}), known as
dissipative soliton \cite{Akhmediev08}, which is the submarine
analog of a terrestrial fairy circle, can be stable
\cite{Woods99,Coullet00}. Many of the bare circles visible in
the coasts of Mallorca and the Adriatic Sea (Fig.
~\ref{fig:PollensaAdriatic}) can be identified with dissipative
solitons.

Thus, our model is able to reproduce and to facilitate
understanding of the nonlinear behavior, leading to the
emergence of dynamic complex landscape patterns in seagrass
meadows. The spatial scale of real patterns is mainly related
to the range of the competing interaction $\sigma_\kappa$. An
estimation of the length scale of the observed patterns can be
obtained using the Fourier spectrum (see the Supplementary
Materials) of images such as Fig. \ref{fig:PollensaAdriatic}a).
It is not always possible to obtain a precise estimate of this
length scale, because often, patterns are not regular enough,
although local hexagonal ordering can be appreciated in the
more regular regions. In these regions, one can identify in the
Fourier spectrum a typical periodicity of $62.9\pm12.7$ $m$
(see the Supplementary Materials). Choosing $\sigma_\kappa=
28.5$ $m$, the critical wavelength at the MI threshold (see the
Supplementary Materials) fits this observed characteristic
length, and simultaneously, a typical hole size of about 30 $m$
is also correctly reproduced. This result allows us to
conjecture the existence of a competitive interaction with a
range of around 20 to 30 $m$, whose specific nature is not yet
known. In addition to their local contribution, the competition
for natural resources (for example, dissolved inorganic
nutrients or CO$_2$) and the interactions mediated by toxic
compounds, such as sulfide, which is accumulated in the soil
\cite{borum2014eelgrass}, are expected to contribute to
nonlocal competitive interactions. Other hypotheses include
interaction through hydrodynamics, which may modify the
sedimentary delivery of nutrients.

The time scales for the landscape dynamics captured by the
model are long, encompassing decades to millennia (movies S1 to
S5 and the Supplementary Materials), depending on the
characteristic demographic time scales of the species. For
instance, \textit{P. oceanica} clones are long-lived organisms,
with clones living tens of thousands of years
\cite{arnaud2012implications} and forming meadows over
centuries to millennia
\cite{duarte1995submerged,kendrick2005modelling}, whereas
\textit{C. nodosa} grows much faster and can form meadows over
decades to centuries \cite{duarte1995submerged}. Hence, the
dynamics conducive to formation of complex landscape patterns
and the transition between them are too slow to be observed
empirically and can be grasped only through a modeling
approach, such as the one developed here (movies S1 to S6),
with parameter values properly constrained by present-day
observations.

In addition to providing qualitative understanding of how
demographic imbalances affect the spatial configuration of
seagrass meadows, the model proposed provides a remarkable,
given its parsimony, description of observed patterns in
seagrass meadows (Fig.~\ref{fig:meadows}). Two additional model
components are needed to reproduce observed density patterns: a
decline in mortality rate, $\omega_{d0}$, with the distance
from the coast, $x$, and a spatial random noise term that
mimics irregular spatial variability of the parameters.
Simulations in small systems and under ideal conditions (that
is, in the absence of noise) may yield perfectly periodic
patterns (insets in Fig.~\ref{fig:biffurcation}), whereas
simulations in large domains, including noise, better resemble
observed patterns (Figs. \ref{fig:meadows} and
\ref{fig:cymodocea}, movies S4 to S6, and the Supplementary
Materials).

The increase in area coverage from the shore toward moderate
depths is characteristic of \textit{P. oceanica} meadows (Fig.
~\ref{fig:meadows}c) [as is also the its decrease and
disappearance toward deep waters; \cite{duarte1991seagrass}].
This indicates high mortality rates in shallow, nearshore
waters, which could be attributed to scouring by waves, leading
to the unpopulated model solution in waters shallower than the
upslope limit of the seagrass, and low mortality rates in
moderately deeper waters, allowing the formation of stable
homogeneous meadows. A smooth decline in mortality with depth
should then lead to complex spatial patterns at intermediate
depths, where the homogeneous solution is unstable (see
Fig.~{\ref{fig:biffurcation}). Including such decline in
mortality (Fig.~\ref{fig:meadows}b) with added noise (see the
Supplementary Materials), the resulting patterns
(Fig.~\ref{fig:meadows}a) accurately reproduce observed
features in \textit{P. oceanica} meadows
(Fig.~\ref{fig:meadows}c), such as more elongated vegetation
gaps near the shore and scattered gaps close to the homogeneous
meadow. The transition from bare soil to patterns with
elongated gaps signals that this region experiences a steep
mortality decrease from very high to moderate values where
hexagons begin to be unstable with respect to stripes.
Scattered gaps are well reproduced close to the homogeneous
meadow, in a mortality range where hexagonal gap patterns
coexist with the homogeneous populated state (see Fig.
\ref{fig:biffurcation}) and dissipative solitons (fairy
circles) can form. Further downslope, the homogeneous meadow
prevails.

The model also reproduces the decline of shoot density and its
variability with depth observed in meadows along the littoral
of the Balearic Islands (Fig. \ref{fig:meadows}h). Shoot
density was measured by scuba divers at random positions in the
meadows without previous knowledge of their spatial
distribution. The results consistently showed low shoot density
variability at depths $>$ 10 $m$ compared to high variability
at shallower depths ($<$ 10 $m$), ranging from close to 0 to
2000 $shoots/m^{2}$, a variability much larger than that in
deeper regions (thick blue dots Fig. \ref{fig:meadows}h). Our
model suggests that high shoot density variability in shallow
waters is a consequence of the presence of complex spatial
patterns near the coast. Simulations using noisy and
depth-dependent mortality and carrying capacity, $b(x,y)$ (see
Fig. \ref{fig:meadows}g), account for the decrease in shoot
density with depth and generate patterns of shoot density that
capture the dispersion of the density close to the coast where
patterns form (Fig. \ref{fig:meadows} e to h, and movie S5).
The model also reproduces complex patterns in meadows of
\textit{C. nodosa}, such as the transition from holes to
patches observed in one of the meadows (Figs.
\ref{fig:PollensaAdriatic}, and \ref{fig:cymodocea}c). Because
this transition occurs parallel to the coast, that is, at a
uniform depth, we inferred this pattern to be derived from a
sudden increase in the mortality rate along the shore (see Fig.
\ref{fig:cymodocea}b). The resulting simulated pattern (Fig.
\ref{fig:cymodocea}a) reproduces very well the observed
features of the real meadow, further confirming that the
complex seascape of fairy circle patterns observed in
Mediterranean seagrass meadows can be reproduced parsimoniously
as a consequence of variability in the seagrass demographic
balance caused by spatial interaction and nonlinearity.

\section*{DISCUSSION}
We have demonstrated that complex landscape patterns, such as
those dotted by fairy circles characteristic of arid
grasslands, are also common features of seagrass seascapes in
the Mediterranean and have also been reported in seagrass
meadows elsewhere
\cite{borum2014eelgrass,den1971dynamic,frederiksen2004spatial,hemminga2000seagrass}.
The parsimonious model developed here demonstrates that fairy
circle seascapes emerge as consequences of nonlinearity and
spatial interactions in seagrass meadows at critical levels of
demographic imbalances, typically met in relatively shallow
nearshore areas of seagrass meadows. In contrast, comparatively
low mortality rates in deeper areas lead to a prevalence of
stable continuous meadows toward the deeper ranges of seagrass
meadows. The model developed here, based on simple inherent
growth traits of the seagrass species and variable mortality
due to nonlocal interaction and nonlinearity, is able to
reproduce the range of complex landscape configurations,
including striped, hexagon, and soliton- dominated landscapes
encountered between the bare sediments and continuous meadow
end members for these landscapes. The model results are robust
enough as to allow inferences on the demographic status of the
meadows on the basis of observed landscape configurations. In
particular, positive hexagons signal the proximity of tipping
points where further increase in seagrass mortality relative to
growth may lead to catastrophic loss of seagrass meadows
\cite{Rietkerk2004,siteur2014beyond}. Because seagrass
ecosystems rank among the most threatened ecosystems globally
\cite{waycott2009accelerating}, the capacity to diagnose the
proximity of seagrass meadows to tipping points for
catastrophic loss based on landscape configurations provides a
tool to guide conservation measures aimed at preventing further
losses.

\section*{MATERIALS AND METHODS}

\subsection*{Derivation of the Advection-Branching-Death model}

Focusing on the three main mechanisms involved in the growth of
clonal plants, namely, apices' linear growth, branching, and
death, we developed a set of partial differential equations
(PDEs; or, more precisely, integro-differential equations) for
the density of shoots and apices. These mechanisms were
identified and implemented in a previous model, which focused
on individual shoots \cite{sintes2005,sintes2006modeling}.
Here, we formulated these mechanisms in terms of upscaled
continuous densities, thus allowing for the description of much
larger spatial and temporal scales. The spatial density of
shoots at position $\vec{r}=(x,y)$ of the sea bottom at time
$t$ is $n_s(\vec{r},t)$, and the density of apices growing in
the direction given by the angle $\phi$ is
$n_a(\vec{r},\phi,t)$. Note that the magnitude of the apices'
growth velocity $\nu$ is assumed constant, and a growth
velocity vector can be written as $\vec{v}(\phi)=(\nu \cos
\phi,\nu \sin \phi )$. The total density of shoots, $n_t$,
considering for simplicity that an apex is carrying a shoot, is
the sum of shoots $n_s$ and apices $n_a$ growing in all
directions,
$n_t(\vec{r},t)=n_s(\vec{r},t)+\int_0^{2\pi}n_a(\vec{r},\phi,t)d\phi$.

Two PDEs describing the evolution of the densities of shoots
and apices can be derived in terms of the contributions of the
three growth mechanisms to the number of shoots in an
infinitesimal surface. First, the number of apices growing in
direction $\phi$ at $t+dt$ in an infinitesimal surface of area
$dxdy$ located at $\vec{r}$ will be the sum of two
contributions: (i) the apices that remain alive coming from
$\vec{r}-\vec{v}(\phi)dt$ because of rhizome elongation and
(ii) new apices that appear because of branching from
directions of growth $\phi+\phi_b$ and $\phi-\phi_b$, which are
the only directions contributing to the growth in direction
$\phi$. $\phi_b$ is the branching angle. Note that those apices
that go away due to rhizome elongation are contributing to
position $\vec{r}+\vec{v}(\phi)dt$. Then, we obtain
\begin{eqnarray}
\label{E1}
n_a(\vec{r},\phi,t+dt)dxdy = (1-\omega_d dt) n_a(\vec{r}-\vec{v}dt,\phi,t)dxdy \nonumber \\
+ \frac{\omega_b dt}{2}\left(n_a(\vec{r},\phi+\phi_b,t)
+n_a(\vec{r},\phi-\phi_b,t)\right)dxdy,
\end{eqnarray}

where $\omega_b$ and $\omega_d$ are branching and death rates,
respectively. Making a Taylor expansion of Eq. \ref{E1} and
neglecting second-order terms and higher, we obtain
\begin{eqnarray}
\label{E2}
n_a(\vec{r},\phi,t) + \partial_t n_a(\vec{r},\phi,t)dt= \nonumber \\
(1-\omega_d dt) n_a(\vec{r},\phi,t) -\vec{v}dt\cdot\vec{\nabla}
n_a(\vec{r},\phi,t) \nonumber \\
+ \frac{\omega_b dt}{2}\left(n_a(\vec{r},\phi+\phi_b,t) +n_a(\vec{r},\phi-\phi_b,t)\right)
\end{eqnarray}

Rewriting Eq. \ref{E2}, we obtain the PDE that describes the
growth of apices in the direction $\phi$
\begin{eqnarray}
\label{E3}
\partial_t n_a(\vec{r},\phi,t) = -\omega_d n_a(\vec{r},\phi,t)
-\vec{v}(\phi)\cdot\vec{\nabla} n_a(\vec{r},\phi,t) \nonumber \\
		+ \frac{\omega_b}{2}\left(n_a(\vec{r},\phi+\phi_b,t)
+n_a(\vec{r},\phi-\phi_b,t)\right),
\end{eqnarray}

where $\vec{\nabla}=(\partial_x,\partial_y)$.\\
The same procedure can be used to obtain the equation for the
shoot density. The first contribution is the shoots that remain
alive at the same position, and the second contribution is due
to the apices that survive and go away in any direction leaving
a shoot behind
\begin{eqnarray}
\label{E4}
n_s(\vec{r},t+dt)dxdy=(1-\omega_d dt)n_s(\vec{r},t)dxdy \nonumber \\
+\frac{\nu}{\rho}dt(1-\omega_d dt)\int_0^{2\pi} n_a(\vec{r},\phi,t)dxdy d\phi,
\end{eqnarray}

where $\rho$ is the distance between shoots.  Using the Taylor
expansion and keeping first-order terms only, we have
\begin{equation}
\label{E5}
n_s(\vec{r},t)+\partial_t n_s(\vec{r},t)dt =(1-\omega_d
dt)n_s(\vec{r},t)+\frac{\nu}{\rho}dt\int_0^{2\pi} n_a(\vec{r},\phi,t)d\phi,
\end{equation}

 which leads to the PDE for the shoot population density
 \begin{equation}
 \label{E6}
 \partial_t n_s(\vec{r},t) = -\omega_d n_s(\vec{r},t) + \frac{\nu}{\rho}\int_0^{2\pi} n_a(\vec{r},\phi,t) d\phi.
 \end{equation}

 The time evolution of a meadow can then be described by two coupled PDEs,
 Eqs. \ref{E3} and \ref{E6}, one for each population density. We name this set of
 two equations the ABD model. The first term in Eqs. \ref{E3} and \ref{E6} corresponds
 to death of shoots and apices. The same death rate $\omega_d$ is considered.
 The second term in Eq. \ref{E3} is an advection in the direction of the elongation
 $\vec{v}(\phi)$ of the rhizome, describing the movement of the apices. The last term
 in Eq. \ref{E3} corresponds to the branching process. Finally, the second
 term in Eq. \ref{E6} accounts for the shoots left behind by the apices. The death rate is given by
 \begin{equation}
 \label{E8}
 \omega_d[n_t(\vec{r},t)]= \omega_{d0} + b n^2_t +\int \int
 \mathcal{K}(\vec{r}-\vec{r}')(1-e^{-a n_t(\vec{r}')}) d\vec{r}' ,
 \end{equation}

 \begin{equation}
 \label{E9}
 \mathcal{K}(\vec{r})=\kappa
 \mathcal{G}(\sigma_\kappa,\vec{r})-\mu\mathcal{G}(\sigma_\mu,\vec{r}).
 \end{equation}

where $\mu=\omega_{d0}$ for simplicity and $\omega_{d0} > 0$.
Both  interaction terms in Eq. \ref{E9} are considered to have
a Gaussian shape
$\mathcal{G}(\sigma,\vec{r})=e^{-r^2/(2\sigma^2)}/(\sigma^2
2\pi)$, where $r^2=x^2+y^2$. Other kernels have been considered
in the literature in different contexts \cite{Pigolotti10},
although qualitatively, the pattern formation feature does not
depend strongly on the precise shape of the kernel
\cite{Colet2014,Gelens2014}, provided it decays faster or equal
than exponential \cite{fernandez2013strong}. Figure S5 shows
the form of the kernel used in this work. Thus, the interaction
is stronger for short distances and decreases exponentially
fast with $r^2$. Because $\mathcal{G}$ is normalized to 1, if
$\kappa > \omega_{d0}$ ($\kappa < \omega_{d0}$), then the
interaction is overall competitive (facilitative). Depending on
the values of $\sigma_\kappa$ and $\sigma_\mu$, competition or
facilitation may dominate at short or long distances. The term
$(1-e^{-an_t})$ be expanded for low densities as $(1-e^{-an_t})
\simeq an_t$, leading to the usual nonlocal term in
Lotka-Volterra\^{a}��like models \cite{Pigolotti07}. The
exponential has been introduced to saturate the interaction
strength for high densities, such that the mortality rate
$\omega_d$ cannot become negative because of the facilitative
interaction leading to the local creation of plants, which is
unreal because a new shoot can be created only through the
growth of the rhizomes or a branching event. For low densities,
then, parameter $a$ multiplies the strength of the nonlocal
interaction. However, the larger the parameter $a$, the faster
the saturation of the interaction as the density grows. Varying
$a$ and $\kappa$, one can change the relative strength between
competition and facilitation. Here, we chose a kernel that is
facilitative at short ranges and competitive at larger scales
(Fig. S5).

%\bibliographystyle{ScienceAdvances}
%\bibliography{Ref}

\section*{Acknowledgements:}
\begin{itemize}
\item We acknowledge helpful discussions with C. L\'opez.
    Funding: D.R.-R., D.G., T.S., E.H.-G., and N.M.
    acknowledge financial support from AEI/FEDER [Agencia
    Estatal de Investigacio\`{I}�n/Fondo Europeo de Desarrollo
    Regional, European Union (EU)] (FIS2015-63628-C2-1-R,
    FIS2015-63628-C2-2-R, and CGL2015-71809-P). C.M.D. was
    supported by King Abdullah University of Science and
    Technology through the baseline funding.
\item Author contributions: D.G., E.H.-G., and T.S.
    conceived and designed the research. D.R.-R., D.G.,
    E.H.-G., and T.S. derived the ABD model. D.R.-R.
    performed the numerical simulations supervised by D.G..
    N.M. and C.M.D. provided the experimental data. D.R.-R.
    processed and analyzed the experimental data. All
    authors contributed to scientific discussions and to
    the writing of the paper.
\item Competing interests: The authors declare that they
    have no competing interests.
\item Data and materials availability: All data needed to
    evaluate the conclusions in the paper are present in
    the paper and/or the Supplementary Materials.
    Additional data related to this paper may be requested
    from the authors.
\end{itemize}

\newpage
\begin{figure}[H]
	\includegraphics[width=0.52\textwidth]{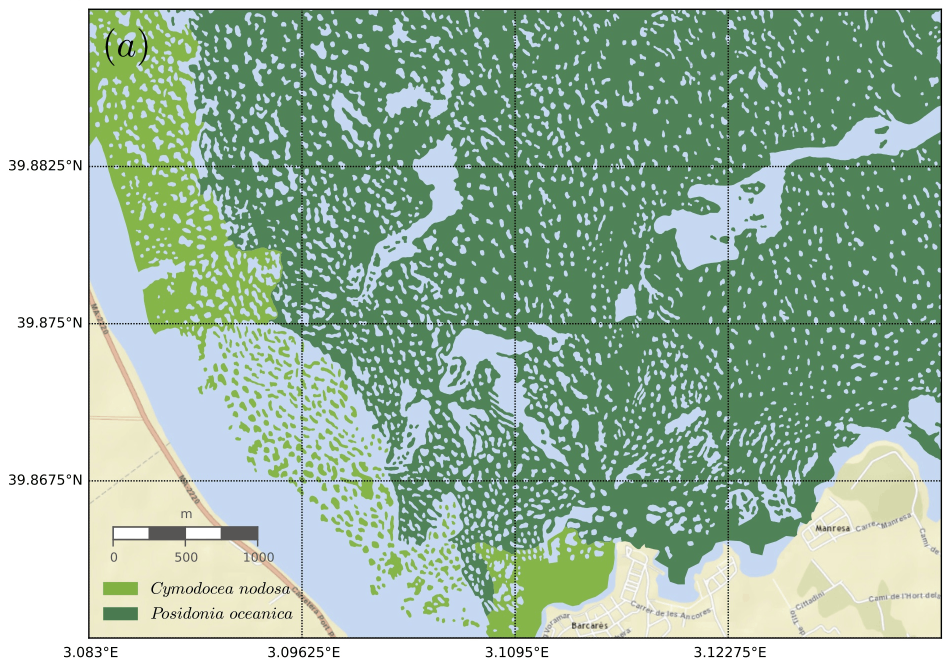}
	\includegraphics[width=0.45\textwidth]{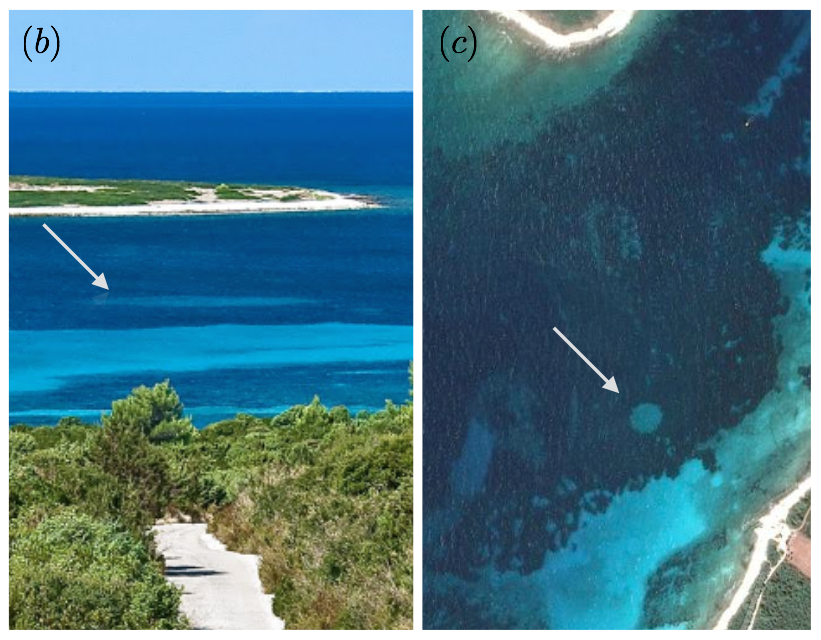}
	\caption{Examples of fairy circles and spatial patterns in Mediterranean seagrass meadows. (a) Side-scan image of a seagrass meadow in Pollen\c{c}a bay (Mallorca Island, Western Mediterranean) from LIFE Posidonia \cite{LIFE_Posidonia} showing different patterns in meadows of \textit{P. oceanica} and \textit{C. nodosa}. Other examples are shown in Figs. S9-S11 in the Suplementary Material. (b) Image of a fairy circle in a \textit{P. oceanica} meadow in the Adriatic Sea as seen from the coast. Phototgraphy by Zvaqan available in Google Street View and (c) the same fairy circle in a satellite image of Google maps (44$^\circ$ 05'37.5"N 14$^\circ$ 55'37.6"E). Other fairy circles can be found at the following locations: 44$^\circ$ 04'01.8"N 14$^\circ$ 57'53.3"E, 39$^\circ$ 08'48.2"N 2$^\circ$ 56'07.1"E.}
\label{fig:PollensaAdriatic}
\end{figure}

\newpage
\begin{figure}[H]
	\includegraphics[width=\columnwidth]{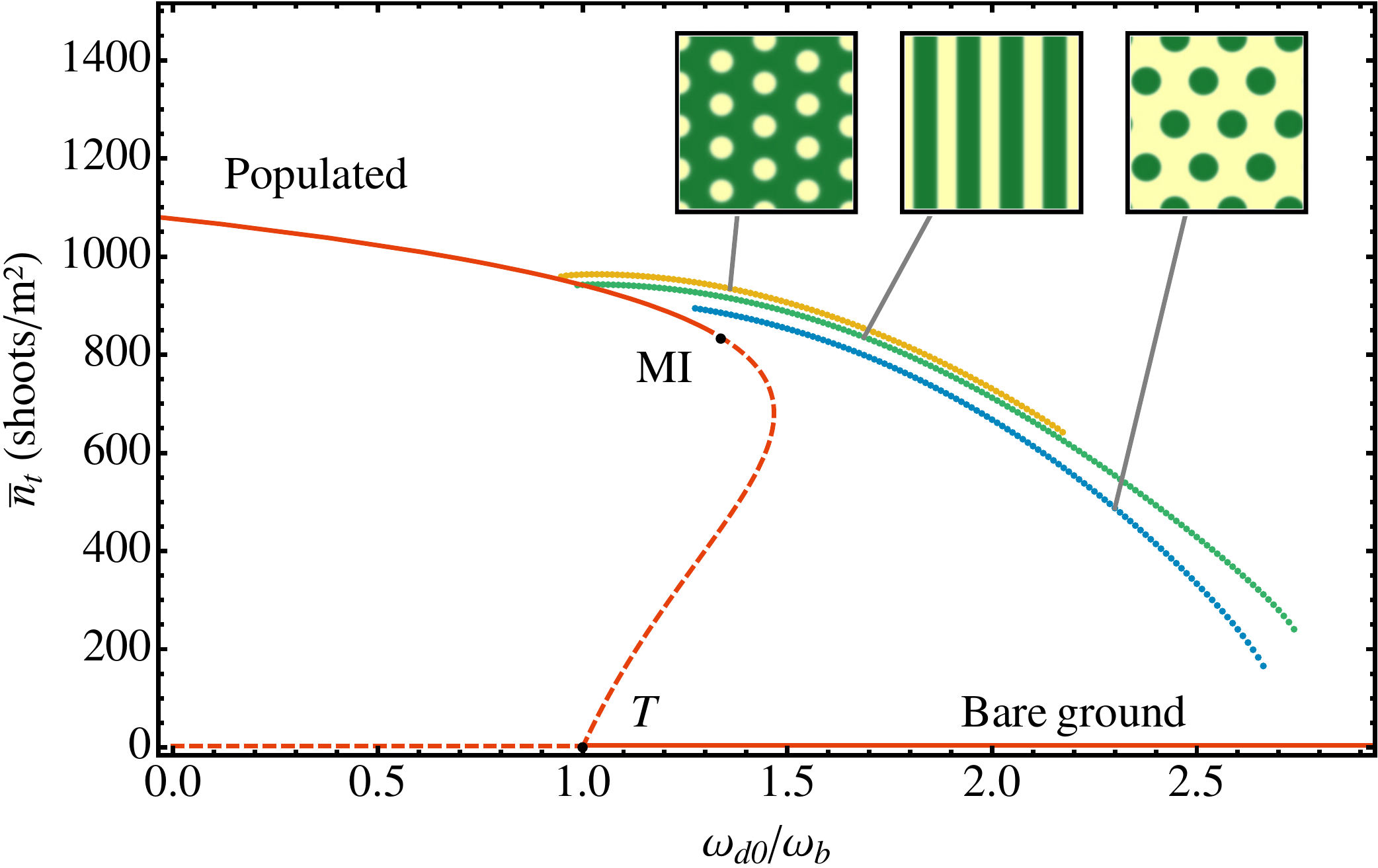}
\caption{Mean shoot density $\bar{n}_t$ (that is,  total number of shoots divided by the whole simulation area) as a function of normalized mortality $\omega_{d0}/\omega_b$ for five different solutions of the ABD model for homogeneous $\omega_{d0}$ and homogeneous $b$. Homogeneous populated and unpopulated states (red), hexagonal arrangement of fairy circles in yellow, stripes in green, and hexagonal arrangement of spots in blue. Solid (dashed) lines indicate stable (unstable) solutions. The insets show the vegetation patterns in the inhomogeneous cases. Only the stable part of the pattern branches is shown, as obtained from direct numerical simulations of the model. MI corresponds to the modulational instability of the populated state, and T corresponds to the transcritical bifurcation of the bare soil. We take parameter values in a range that reproduce patterns seen in side scans, using typical values for \textit{P. oceanica} for the parameters already known [see the study by Sintes et al. \cite{sintes2006modeling} and references therein]: $\omega_b=0.06~year^{-1}$, $\nu=6.11~cm/year$, $\rho=2.87~cm$, $\phi_b=45^\circ$, $b=1.25~cm^4 year^{-1}$, $\kappa=0.048~year^{-1}$, $\sigma_\kappa=2851.4~cm$, $a=27.38~cm^2$, $\sigma_\mu=203.7~cm$, $\mu=\omega_{d0}$ (see the Supplementary Materials). The formation of the three patterned solutions (negative hexagons, stripes, and positive hexagons) is shown in movies S1, S2, and S3, respectively.}
\label{fig:biffurcation}
\end{figure}

\newpage
\begin{figure}[H]
 \includegraphics[width=\columnwidth]{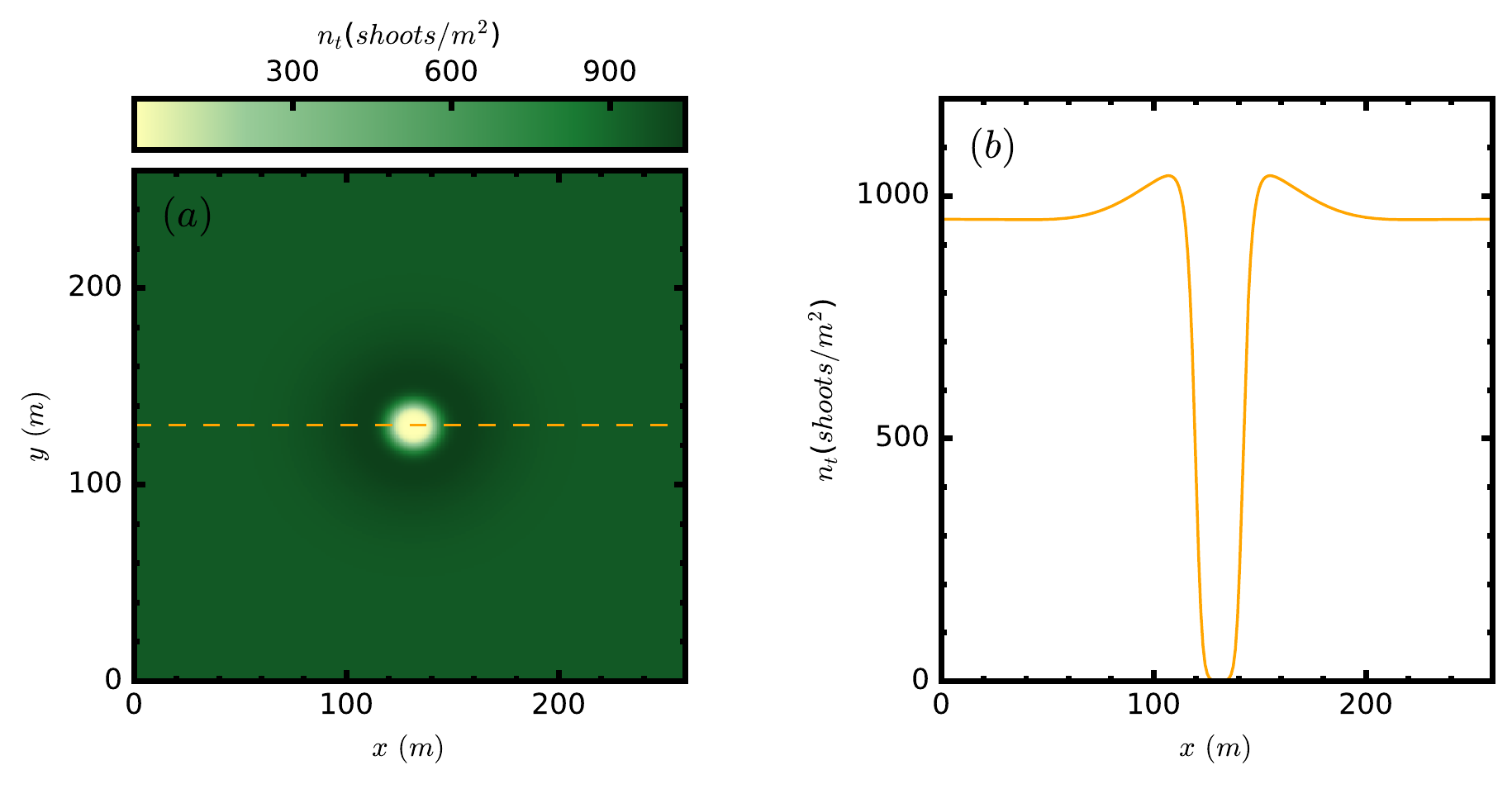}
\caption{Spatial distribution of the shoot density (high densities are represented in dark green and low ones in bright yellow) in a simulation of a \textit{P. oceanica} meadow showing a stable fairy circle. The fairy circle, or dissipative soliton, is clear in the density profile (b) along the transverse cut shown in panel (a) by a dashed line. Here, $\omega_{d0} = 0.057 year^{-1}$. Other parameters are the same as in Fig. \ref{fig:biffurcation}}
\label{fig:hole}
\end{figure}

\newpage
\begin{figure}[H]
	\includegraphics[width=0.5\textwidth]{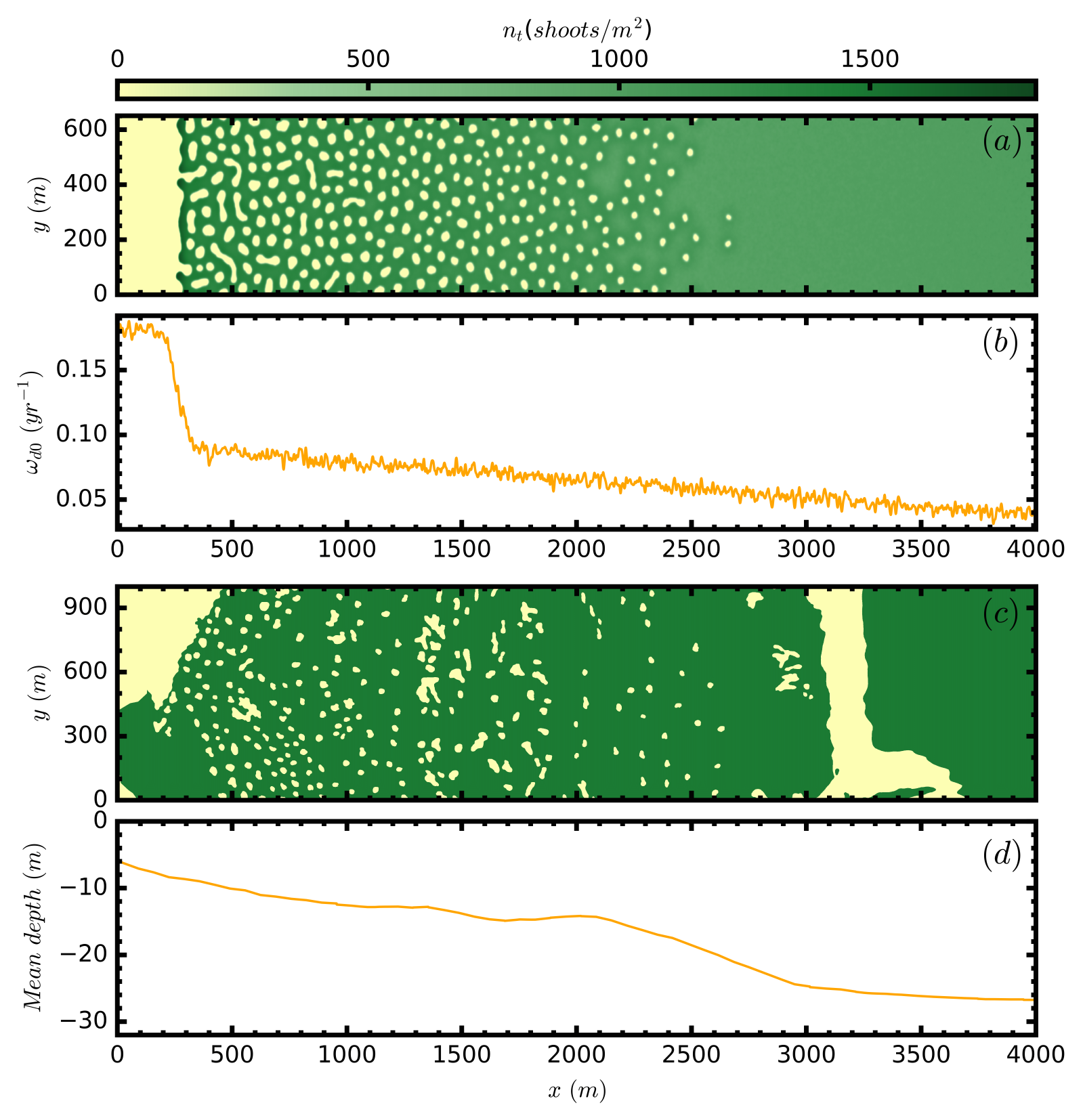}
	\includegraphics[width=0.5\textwidth]{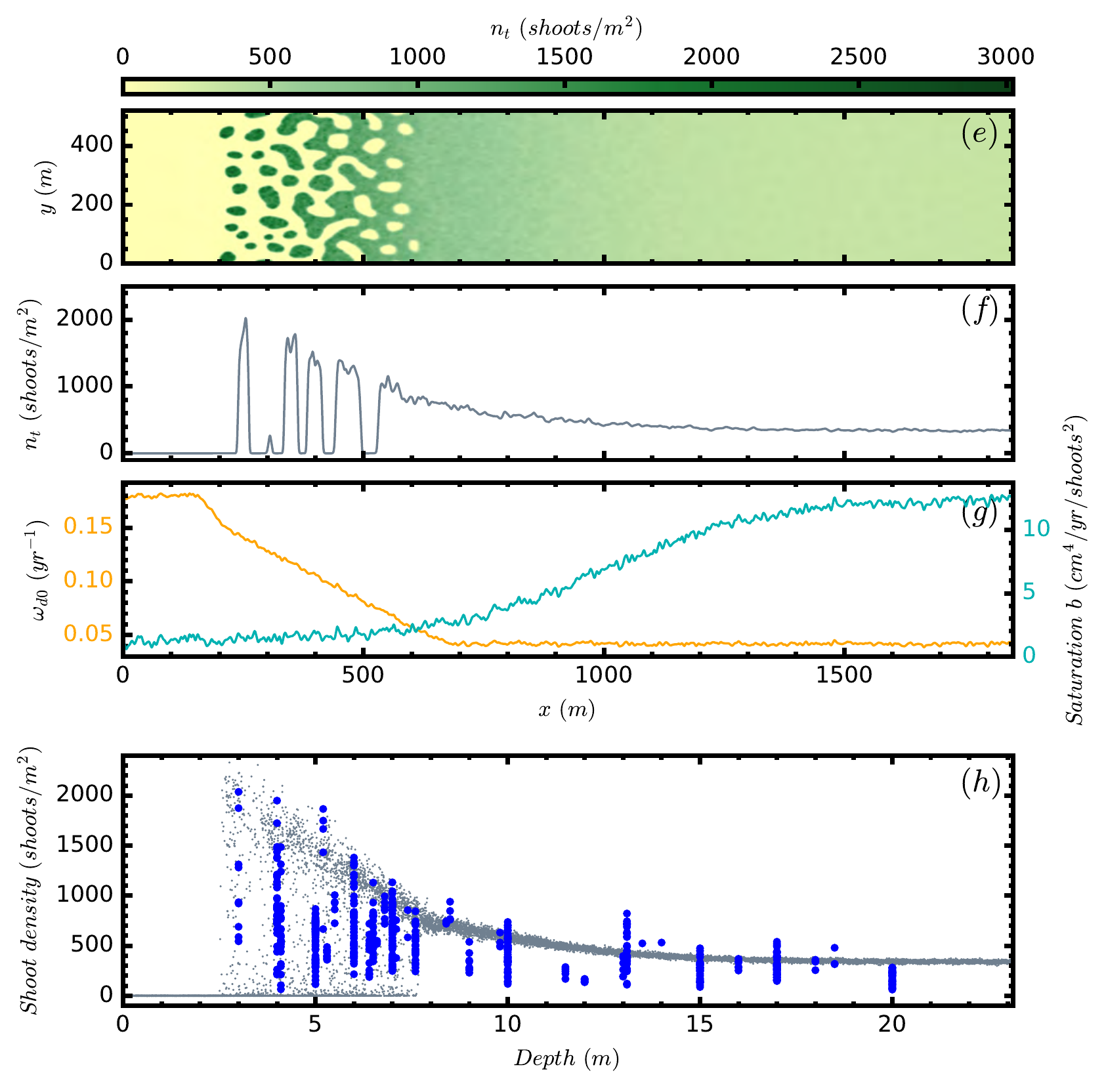}
	\caption{Comparison of numerical simulations with patterns observed in
seagrass meadows. (a) Final spatial density distribution of shoots from a numerical
simulation of the ABD model that uses the mortality profile plotted in panel (b).
(c) Observed coverage \cite{LIFE_Posidonia} of \textit{P. oceanica} from LIFE Posidonia
side-scan cartography in the Balearic coast area limited by the following coordinates:
39$^\circ$ 45'54.1"N 3$^\circ$ 09'49.5 E; 39$^\circ$ 47'25.6"N 3$^\circ$ 11'48.7 E;
39$^\circ$ 47'48.6"N 3$^\circ$ 11'19.0 E; and 39$^\circ$ 46'17.1"N 3$^\circ$ 09'19.9 E.
(d) Depth in that region averaged along the y direction. (e to h) Comparison of a numerical
simulation with field density measures: (e) Spatial density distribution of \textit{P. oceanica}
as obtained from numerical simulations with a custom spatially dependent mortality
[orange line in (g)), left scale] and a profile of the saturation strength $b(x,y)$
[blue line in (g), right scale]. (f) Cut of (e) at $y=102$. (h) Observed \textit{P. oceanica}
density measured by scuba divers (in blue, data file S1) as function of the depth for
different locations spread over the coastline of the Balearic Islands and the density
in random locations of the numerical simulation shown in (e) (gray). Parameters are
same as in Fig.~\ref{fig:biffurcation}. The time evolutions of the simulations are
shown in movies S4 and S5.}
\label{fig:meadows}
\end{figure}

\newpage
\begin{figure}[H]
	\includegraphics[width=\columnwidth]{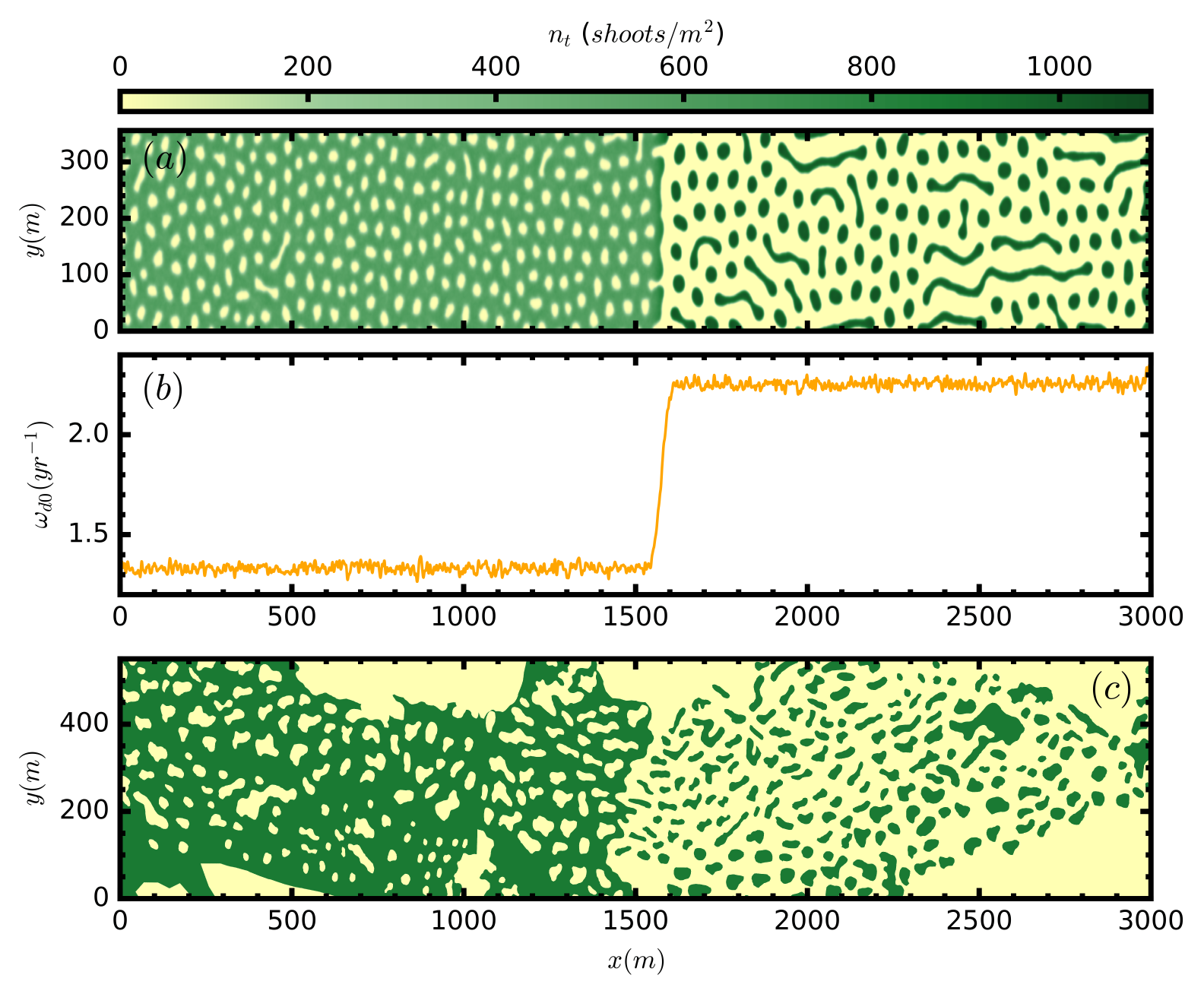}
\caption{Comparison of numerical simulation with patterns
observed by side-scan sonar \cite{LIFE_Posidonia} for a region of
coexistence between holes and patches in a meadow of \textit{C. nodosa}
in Mallorca Island (Fig.\ref{fig:PollensaAdriatic}). The set of model parameters
for \textit{Cymodocea nodosa} is $\omega_b=2.3$ $year^{-1}$, $\nu=160$ $cm/year$,
$\rho=3.7$ $ cm$, $\phi_b=45^\circ $, $b=112.71$ $cm^4 year^{-1}$,
$\kappa=2.76$ $year^{-1}$, $\sigma_\kappa=2226.1$ $cm$, $a= 21.0$ $ cm^2$, $\sigma_\mu=139.1$ $ cm$,
$\mu=\omega_{d0}$, and the area modeled (a subset of that shown in Fig. \ref{fig:PollensaAdriatic}a)
is bounded by the coordinates: 39$^\circ$ 53'16.4"N 3$^\circ$ 05'12.7 E;
39$^\circ$ 51'52.0"N 3$^\circ$ 06'15.7 E; 39$^\circ$ 51'43.1"N 3$^\circ$ 05'55.6 E;
39$^\circ$ 53'07.5"N 3$^\circ$ 04'52.6 E (movie S6). (a) Final spatial density
distribution of shoots from a numerical simulation of the ABD model using the mortality
profile shown in (b). (c) Observed coverage \cite{LIFE_Posidonia} of \textit{C. nodosa}
from LIFE Posidonia side-scan cartography in the Balearic coast.}
\label{fig:cymodocea}
\end{figure}

\section*{SUPPLEMETNARY MATERIALS}
\section*{MATERIALS AND METHODS}

\subsection*{Homogeneous solutions and linear stability analysis}
To simplify the calculations it is convenient to work with dimensionless units,
such that time, space and density of shoots and apices in the new units are given by:
$T=\omega_b t$, $\vec{R}=\frac{\omega_b}{\nu} \vec{r}$, $n'_s= \sqrt{\frac{b}{\omega_b}} n_s$ and
$n'_a= \sqrt{\frac{b}{\omega_b}} n_a$. We note that the branching rate fixes the
temporal scale, the spatial scale is determined by the velocity of the rhizome elongation,
and the scale of the number of shoots is determined by the saturation parameter $b$.
In the following we drop the primes from the variables and parameters expressed in the new units.

The homogeneous stationary solutions can be obtained from Eqs.
\ref{E3} and \ref{E6} on the main text by setting the temporal
and spatial derivatives equal to zero. Two solutions can be
found, the trivial zero solution, which we will refer to as the
unpopulated solution, and the populated solution, given by
following implicit equation:
\begin{eqnarray}
\label{E11}
n_t^{*2}+\omega_{d0}e^{-a n_t^{*}} -1 +\kappa (1-e^{-a n_t^{*}})= 0 \nonumber
\\
n_s^{*}= \frac{1}{1+\rho}n_{t}^{*} \nonumber \\
N_a^{*}=\rho n_s^{*}
\end{eqnarray}

where $N_a=\int_0^{2\pi}n_a(\phi)d\phi$. Here we consider that
the density of apices is the same for all the growing
directions, and therefore $N^*_a=2\pi n_a^*$. We focus in the
positive-density solutions. Mathematical solutions with
negative density exist, but we neglect them as they have no
biological meaning. We are mainly interested in the dependence
of the populated solution on $\omega_{d0}$, that controls the
external stress which the plant is exposed to. Therefore, in
the following, we consider $\omega_{d0}$ as the main control
parameter. The dependence of the populated solution on
$\omega_{d0}$ is shown in the bifurcation diagrams, where the
homogeneous solution is represented in red (see Fig. 2). We can
distinguish two different regimes depending on the value of the
ratio $\kappa/\omega_{d0}$. When $\kappa/\omega_{d0}<1$, the
interaction is overall cooperative. In this case, the
bifurcation from the unpopulated to the populated states
(occurring at $\omega_{d0}=\omega_b$) is subcritical, and the
populated solution coexist with the unpopulated stated in a
certain range of values of $\omega_b < \omega_{d0} <
\omega_{d,SN_1}$, as can be seen in Fig. 2.  When
$\kappa/\omega_b>1$, the interaction is overall competitive and
there is no coexistence, as can be seen in Fig.
\ref{BifurcationDiagram}. The populated solutions is said to be
supercritical. A priory both scenarios are good candidates to
reproduce the behavior of \textit{P. oceanica}, however the
existence of localized structures is associated to subcritical
parameter sets, and this is the reason why we chose
$\kappa/\omega_b=0.8$ in this work.

To study the stability of the homogeneous solutions we consider
perturbations of the form $n_s=n_s^*+n_{sp}$,
$n_a=n_a^*+n_{ap}$. The linearized systems reads:
\begin{eqnarray}
\partial_T n_{ap}=-\left[\omega_{d0}+(\kappa-\omega_{d0})(1-e^{-a n_t^{*}})+
n_t^{*2}\right] n_{ap} \nonumber \\
+\left[-2n_t^{*}n_{tp}-a e^{-an_t^{*}}\int \int \mathcal{K}(\vec{R}
-\vec{R}')n_{tp}(\vec{R}') d\vec{R} \right]\frac{\rho}{2\pi(1+\rho)}n_t^{*}
\nonumber \\
-\hat{v}(\phi)\cdot\vec{\nabla} n_{ap}+ \frac{1}{2}\left[n_{ap}(\phi+\phi_b)
+n_{ap}(\phi-\phi_b)\right]
\label{E13}
\\
\partial_T n_{sp} = -\left[\omega_{d0}+(\kappa-\omega_{d0})(1-e^{-a n_t^{*}})
+n_t^{*2}\right] n_{sp}  \nonumber \\
+ \left[-2n_t^{*}n_{tp}-a e^{-an_t^{*}}\int \int \mathcal{K}(\vec{R}
-\vec{R}')n_{tp}(\vec{R}') d\vec{R} \right] \frac{n_t^{*}}{(1+\rho)} \nonumber\\
+ \frac{1}{\rho}\int n_{ap}(\phi)d\phi,
\label{E14}
\end{eqnarray}

where $n_{tp}=n_{sp}+\int_0^{2\pi}n_{ap}(\vec{R},\phi,t)d\phi$
and $\hat{v}(\phi)$ is a unit vector in direction $\phi$. Since
the advection term is periodic in $\phi$ [$\hat{v}(\phi)=(\cos
\phi, \sin \phi)$], Eqs. \ref{E13}-\ref{E14} are a set of
linear differential equations with periodic coefficients of
periodicity $2\pi$. Because the dependence in $\phi$ should be
periodic, perturbations can be written in the following form:
\begin{equation}
\label{E15}
n_{ap} (\vec{R},\phi,t)= \sum_{q_{\phi}}
\int\int\tilde{n}_{ap,q_{\phi}}(\vec{q},t) e^{ i \vec{q}\cdot\vec{R}} e^{ i
q_{\phi}\phi}d\vec{q}
\end{equation}
\begin{equation}
\label{E16}
n_{sp} (\vec{R},t)= \int\int\tilde{n}_{sp}(\vec{q},t) e^{ i
\vec{q}\cdot\vec{R}}
d\vec{q},
\end{equation}

where $i$ is the imaginary unit, and
$q_\phi=\dots,-1,0,1,2\dots$. Introducing Eqs.
\ref{E15}-\ref{E16} in \ref{E13}-\ref{E14} we obtain the
following set of coupled linear ordinary differential equations
for the components $\tilde{n}_{ap,q_{\phi}}$, $\tilde{n}_{sp}$
:
\begin{eqnarray}
\partial_T
\tilde{n}_{ap,q_{\phi}} =-\left[\omega_{d0}+(\kappa-\omega_{d0})(1-e^{-a
n_t^{*}})+n_t^{*2}\right]\tilde{n}_{ap,q_{\phi}} \nonumber \\
-\frac{iq_{+}}{2}n_{ap,q_{\phi}+1}-\frac{iq_{-}}{2}n_{ap,q_{\phi}-1}+
\cos(q_{\phi}\phi_b)\tilde{n}_{ap,q_{\phi}}
 \label{E17}
\end{eqnarray}
\begin{eqnarray}
\partial_T \tilde{n}_{ap,0}=\left[1-\omega_{d0}+(\kappa-\omega_{d0})(1-e^{-a
n_t^{*}})+n_t^{*2}\right]\tilde{n}_{ap,0} \nonumber \\
+\left[-2n_t^{*}-a e^{-an_t^{*}} \tilde{\mathcal{K}}(\vec{q})
\right]\frac{\rho}{2\pi(1+\rho)}n_t^{*}\tilde{n}_{tp} \nonumber \\
-\frac{iq_{+}}{2}\tilde{n}_{ap,+1}-\frac{iq_{-}}{2}\tilde{n}_{ap,-1}
 \label{E18}
\end{eqnarray}
\begin{eqnarray}
\partial_T \tilde{n}_{sp} = -\left(\omega_{d0}+(\kappa-\omega_{d0})(1-e^{-a
n_t^{*}})+n_t^{*2}\right)\tilde{n}_{sp} \nonumber \\
+ \frac{\left(-2n_t^{*}-a e^{-an_t^{*}}\tilde{\mathcal{K}}(\vec{q})
\right)}{(1+\rho)}n_t^{*}\tilde{n}_{tp} + \frac{2\pi}{\rho}\tilde{n}_{ap,0},
\label{E19}
\end{eqnarray}

where $q_\pm=q_x \pm i q_y$.\\
Eqs. \ref{E17}-\ref{E19} describe the linear evolution of the
perturbation of the homogeneous solutions. The rsh of this
system of equations can be written in a matrix form of infinite
dimension. Truncating the matrix operator at order
$q_{\phi}=\pm 4$ (neglecting contributions with $|q_{\phi}|>4$)
\footnote{This truncation is equivalent to the numerical
discretization of $\phi$ that has been used for the numerical
simulations.} and diagonalizing numerically we find the growth
rate of perturbations with wavenumber $\vec{q}=(q_x,q_y)$. The
diagonalization leads to 10 eigenvalues for each $\vec{q}$. The
solution is stable if all eigenvalues $\lambda_j(q_x,q_y)$ have
negative real part. On the contrary, if the real part of the
eigenvalue for a given wave number $\vec{q}$ becomes positive,
the homogeneous solution becomes unstable to perturbations with
the corresponding spatial periodicity, and a spatial patterns
forms (see Fig. \ref{BifurcationDiagram}).

Figure \ref{Criticalwavenumber} shows the dispersion relation
of the branch of eigenvalues with largest real part. For the
parameter values considered here, the first eigenvalue that
become positive is a real eigenvalue with $|\vec{q}| \sim 0.1$,
the critical wavenumber. This is a pattern-forming instability
known as Turing or modulational instability (MI).

With this procedure we can determine the regions in the
parameter space ($\omega_{d0}$, $\kappa$) where the populated
and unpopulated homogeneous solutions are stable or unstable.
This is summarized in the phase diagram shown in Fig.
\ref{Phasediagram}. The unpopulated solution is stable if the
branching rate is smaller than the mortality rate
$\omega_{d0}/\omega_{b} >1$ (region 2). If
$\omega_{d0}/\omega_{b} <1$ the density of shoots grows
exponentially in the linear regime, and the systems goes to the
populated solution. $\omega_{d0}/\omega_{b} =1$ corresponds to
a transcritical bifurcation indicated by T in Fig.
\ref{Phasediagram}.

The populated solution exist for $\omega_{d0}/\omega_{b} <1$
and it is stable in blue region 1. However, if
$\kappa/\omega_{b}<1$ (interaction overall cooperative) the
population solution extends to mortalities larger than the
branching rate $\omega_{d0}/\omega_{b} >1$, and it coexists
with the unpopulated solution until the saddle-node bifurcation
line indicated as SN$_1$ (shaded region 3). The populated
solution is unstable to periodic patterns in the yellow region
4. Region 4 is delimited by the MI line.

The linear stability analysis allows us to obtain the critical
wavenumber $|\vec{q}_c|$ of the MI, which determines the
typical periodicity of the pattern ($2 \pi/|\vec{q}_c|$)
arising from the bifurcation. Fig. \ref{Wavenumdependence}
shows the dependence of the corresponding wavelength of the
wavenumber with maximum growth rate as function of the range of
the competitive interaction $\sigma_\kappa$. At the MI this
wavenumber coincides with $|\vec{q}_c|$. From these results,
and using the typical periodicity of real patterns ($62.9\pm
12.7 m$), we can find an estimation of the value of
$\sigma_\kappa$ around $20-30m$. Although the exact number
depends on the precise value of $\kappa$, the dependence on
this parameter is small and the wavenumber is mainly determined
by the value of $\sigma_\kappa$.

\subsection*{Numerical simulations}
\paragraph{Pseudospectral method}
The ABD model is a system of two coupled nonlinear
integro-diferential equations with partial derivatives. A
pseudo-spectral method is used to integrate the time evolution
of the model equations \ref{E3} and \ref{E6} on the main text.
The model is effectively three dimensional, two spatial
dimensions ($x$, $y$), and one angular dimension ($\phi$)
corresponding to the direction of growth of the apices. Since
we are mainly interested in the spatial distribution of the
population densities, we use the minimum number of grid points
in $\phi$ space compatible with the branching angle. In this
case, then, we consider angles multiple of $\pi/4=45 ^{\circ}$,
which describes well the branching angle both for \textit{P.
oceanica} as for \textit{C. nodosa}
\cite{sintes2005,sintes2006modeling}. This means that we
describe the apices growing only in eight different directions.
We have checked that increasing the number of directions does
not qualitative changes the results and it considerable
increases the computational needs. In practice we have then
nine two-dimensional fields: one for the density of shoots and
8 for the density of apices growing in each corresponding
direction. The nine fields, that depend on ($x$, $y$), are
coupled through the branching and the total density in the
nonlocal term. We consider a square grid with $N_x$ and $N_y$
grid points and we integrate the time evolution using a
pseudo-spectral method where the linear terms in Fourier space
are integrated exactly, while the nonlinear terms are
integrated using a second-order in time scheme described in
Ref. \cite{montagne1997wound}.

Numerical simulations provide information about the nonlinear
behavior of the system that can not be predicted by the linear
stability analysis. Typical simulations start with the
homogeneous solution with a superimposed small random
perturbation. In stable regions of parameter space
perturbations decay, and the solution remains, while in regions
where the homogeneous solution is unstable, perturbations grow,
and the nonlinear dynamics send the system to a different
stable solution. In the region unstable to patterns, initial
homogeneous solutions develop modulations that grow until a
pattern forms. Different patterns appear depending on the
parameters, as we can see in Fig. 2 and in Fig.
\ref{BifurcationDiagram}. The regions of stability of positive
and negative hexagons and stripes change with $\kappa$, as well
as the prevalence of one over another. To study the stability
of the different spatial patterns changing mortality we have
performed simulations continuing $\omega_{d0}$. We start with
an initial condition of a pattern, we add small white noise,
and we change the mortality a small amount, letting the system
evolve to reach a new stationary state. We use then, the final
state as initial condition for the next parameter step.
Repeating this procedure we can generate the stable branches
shown in the bifurcation diagrams, where the average densities
of each final state is plotted.

\paragraph{Mortality profiles}
In order to introduce a mortality profile in the simulations we
have to take into account two things. First we have to
introduce a matrix with the values of the mortality at each
position. Second, since the pseudospectral method needs
periodic boundary conditions the introduced profile must be
periodic, in the center the profile can have the desired shape
but opposite boundaries must connect smoothly. The easiest way
to produce a profile is to design it using straight lines and
apply a filter to smooth out the corners. For the last step we
use a diffusion operator in Fourier space. We apply the Fourier
transform to our array and we multiply each component $\vec q$
by the diffusion operator, given by $e^{-q^2t}$, where $t$
controls the softness level. After that we anti-transform to
real space. The resulting profile will preserve the initial
qualitative shape, but it will be smooth and periodic.

Including the mortality profile in the simulations, the
corresponding patterns can be seen in Fig. \ref{meadow}.
However, the unreal conditions of a mortality profile changing
smoothly produce ideally circular bare holes that do not
resemble those present in the real meadows shown in Fig.
\ref{meadow} (c). It is clear that some variability have to be
included to better reproduce the empirical observations.

\paragraph{Noise generation}
To account for irregularities of the sea bottom we introduce
variability on top of the mortality profile. We add a noise
$\chi(x,y)$ with a typical spatial scale to $\omega_{d0}$. This
noise is generated using the following expression:
\begin{equation}
\chi(x,y) = \mathcal{F}^{-1}\left\{e^{\frac{-q^2s^2}{2}}e^{- i 2 	\pi u}\right\}
\end{equation}

where $\mathcal{F}^{-1}$ is the inverse Fourier transform, $q$
is the modulus of the wavevector of each Fourier component, and
$u$ is random number between 0 and 1 with a flat probability
distribution. The parameter $s$ controls the typical spatial
scale of noise. The Gaussian shape in Fourier space inhibits
long wavelength contributions, in such a way the noise is
reasonably smooth. In our simulations we take $s=101.83 m$.
Fig. \ref{NWNoise} shows the spatial variability of the noise
that is added to the mortality profile with a certain
amplitude.

\subsection*{Determination of the characteristic length scale of real
patterns in \textit{P. oceanica} meadows \newline} The LIFE
Posidonia project realized a mapping of the Posidonia meadows
as well as other species such as \textit{C. nodosa} by using
side-scan sonar. As a result, high-resolution images
determining the presence/absence of plants are available. These
maps, available in low resolution in
(\href{http://lifeposidonia.caib.es}{http://lifeposidonia.caib.es})
show the spatial distribution of the \textit{P. oceanica} and
\textit{C. nodosa} meadows displaying patterns. Although there
are many irregularities, bare holes have a typical size and are
separated a characteristic distance. In order to determine the
predominant scale of the pattern we analyze the Fourier
transform of different regions. In some of them it is difficult
to identify a typical length scale. However, a peak indicating
a typical length scale is often clearly visible in} the Fourier
transform. Tables \ref{table1} and \ref{table2} report the 11
cases that we have used to compute the average wavelength of
the patterns. The coordinates of two vertexes of the selected
rectangular regions are given in table \ref{table1}, and the
wavenumber of the peak in the Fourier transform is given in
table \ref{table2}. Fig. \ref{Fourier} shows an example of the
Fourier transform of a cartography image corresponding to the
first row of table \ref{table1}. The average wavenumber is
$0.016 (m^{-1})$ corresponding to a typical wavelength of 63
$m$.

\newpage
\begin{figure}[H]
  \centering
  \setcounter{figure}{0}
  \makeatletter
  \renewcommand{\thefigure}{S\@arabic\c@figure}
  \makeatother
 \includegraphics[scale=0.6]{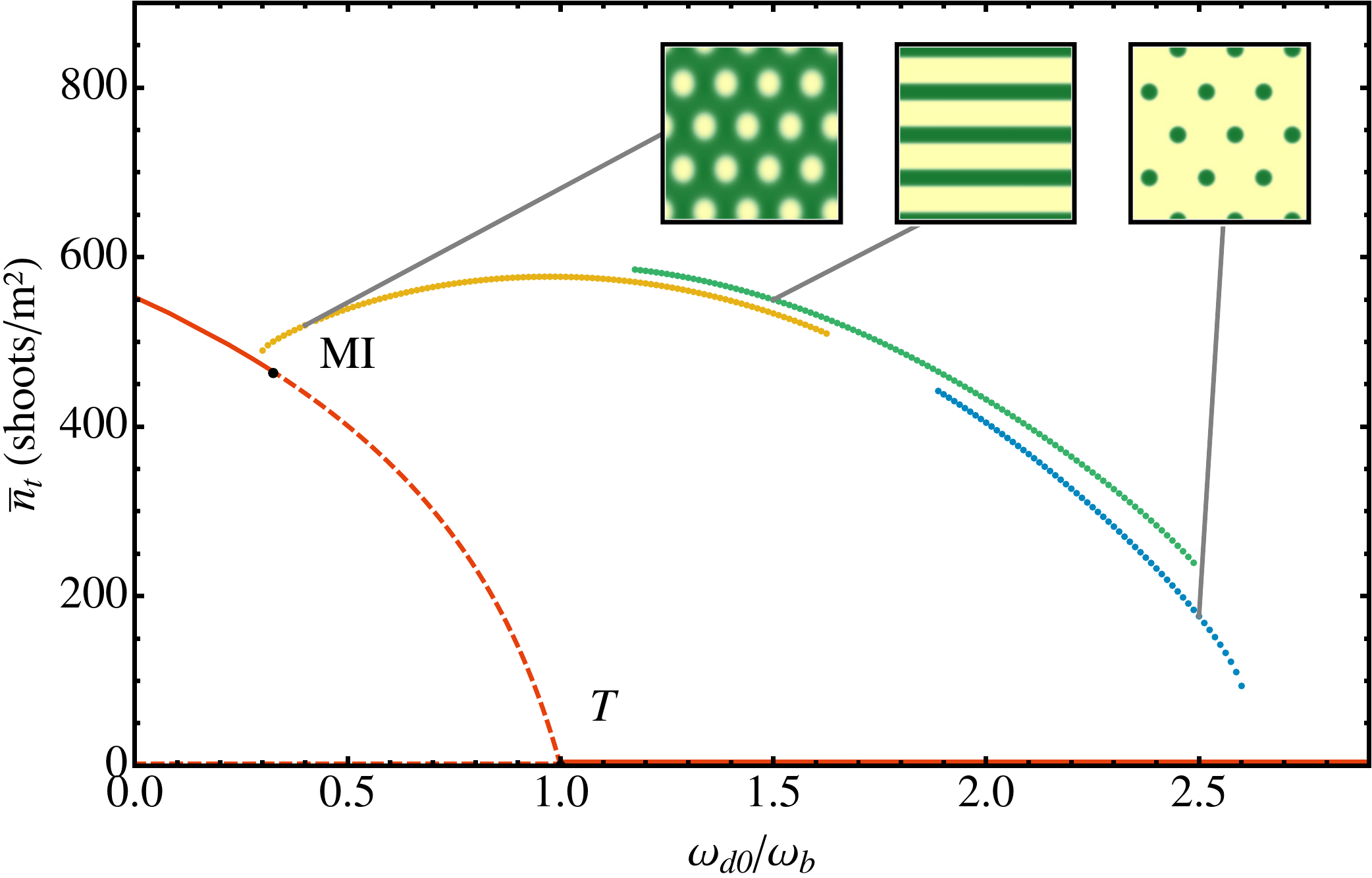}
\caption{Mean shoot density $\bar{n}_t$ (that is, total number
of shoots divided by the whole simulation area) as a function of
normalized mortality $\omega_{d0}/\omega_b$ for the supercritical case.
Five different solutions of the ABD model for homogeneous $\omega_{d0}$
and $b$ are shown: homogeneous populated and unpopulated states (red),
hexagonal arrangement of fairy circles (yellow), stripes (green) and hexagonal
arrangement of spots (blue). Solid (dashed) lines indicate stable (unstable)
solutions. The insets show the vegetation patterns in the inhomogeneous cases.
Only the stable part of the pattern branches are shown, as obtained from direct
numerical simulations of the model. MI corresponds to the modulation instability
of the populated state and T to the transcritical bifurcation of the bare soil.
Here we take parameters for \textit{P. oceanica} as in Fig. 2 of the main text,
but with a higher value of the competitive parameter $\kappa$ so that the T
bifurcation to the populated state is now supercritical. $\omega_b = 0.06$
$ year^{-1}$, $\nu=6.11$ $ cm/year$, $\rho=2.87$ $cm$, $\phi_b=45^\circ$,
$b=1.25$ $ cm^4year^{-1}$, $\kappa = 0.072$ $ year^{-1}$, $\sigma_{\kappa}=$ $2851.4$
$cm$, $a=27.38$ $ cm^2$, $\sigma_\mu=203.7$ $cm$.}
\label{BifurcationDiagram}
\end{figure}

\newpage
\begin{figure}[H]
  \makeatletter
  \renewcommand{\thefigure}{S\@arabic\c@figure}
  \makeatother
\includegraphics[width=1\textwidth]{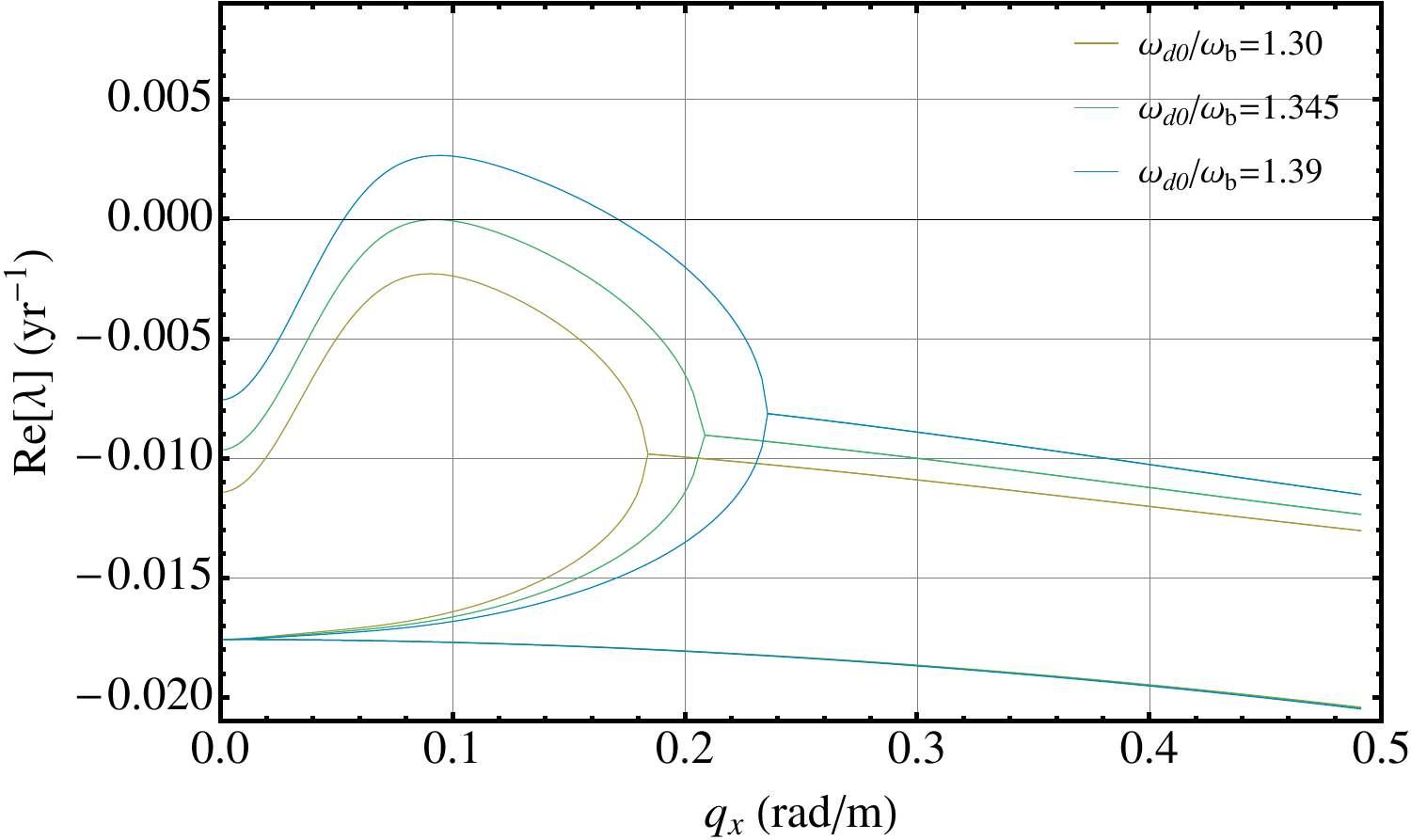}
\caption{Growth rate of perturbation with wavenumber $(q_x,q_y=0)$ close to the MI.
Three different values of the mortality are shown, the yellow curve corresponds to
a stable case, the green one to the critical point, and finally the blue line
corresponds to the unstable case. Here $\omega_b = 0.06$ $year^{-1}$, $\nu=6.11$
$ cm/year$, $\rho=2.87$ $cm$, $\phi_b=45^\circ$, $b=1.25$ $ cm^4year^{-1}$,
$\kappa = 0.048$ $year^{-1}$ , $\sigma_{\kappa}=2851.4$ $cm$, $a=27.38$ $cm^2$,
and $\sigma_\mu=203.7$ $cm$.}
\label{Criticalwavenumber}
\end{figure}

\newpage
\begin{figure}[H]
  \makeatletter
  \renewcommand{\thefigure}{S\@arabic\c@figure}
  \makeatother
 \includegraphics[width=1\textwidth]{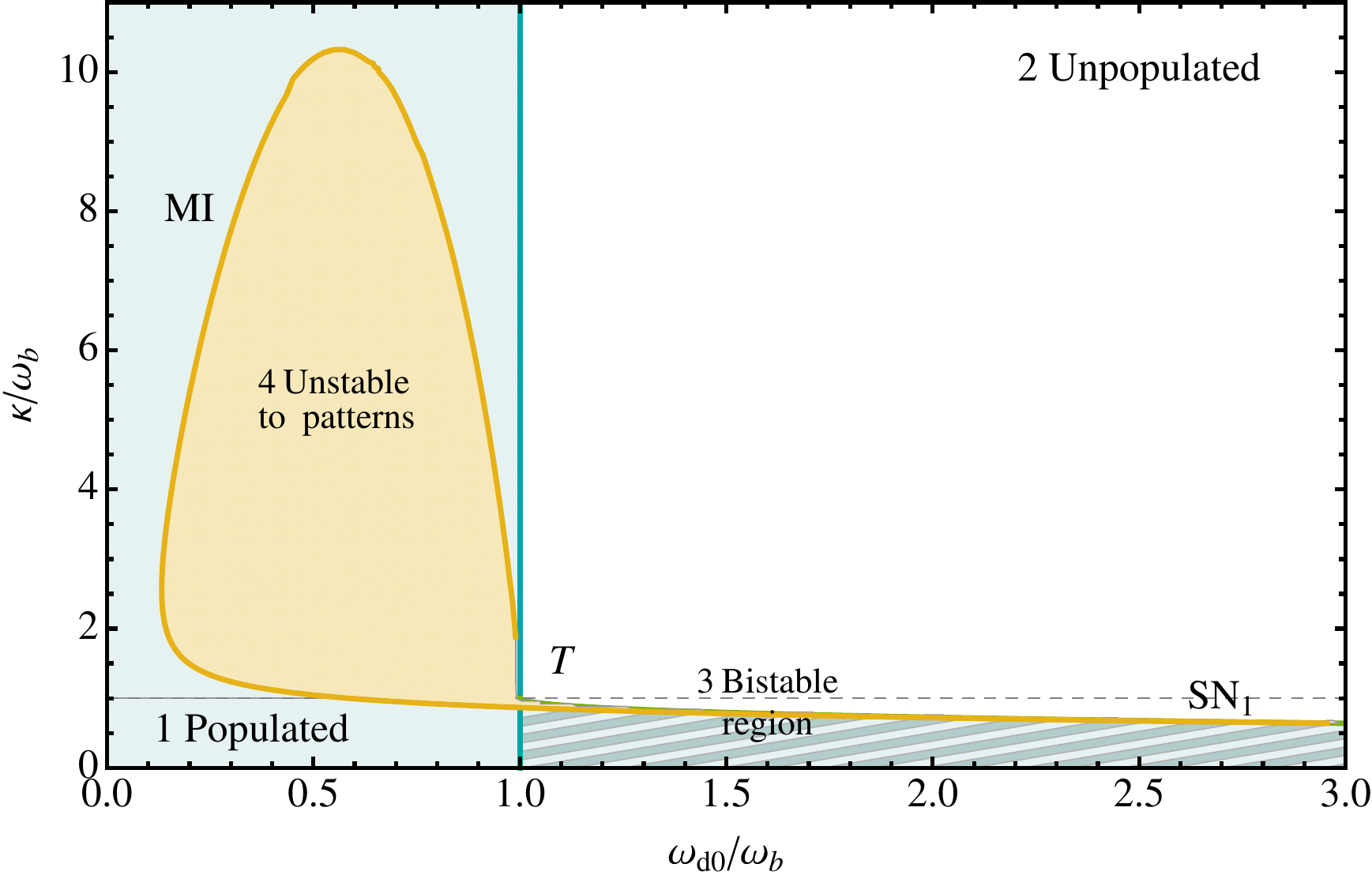}
\caption{Phase diagram of the ABD model for \textit{P. oceanica}. Here
$\omega_b = 0.06$ $ year^{-1}$, $\nu=6.11$ $ cm/year$, $\rho=2.87$ $cm$,
$\phi_b=45^\circ$, $b=1.25$ $ cm^4year^{-1}$, $\sigma_{\kappa}=$ $2851.4$ $cm$,
$a=27.38$ $ cm^2$, $\sigma_\mu=203.7$ $cm$. We represent the region where the
populated solution is stable in blue (region 1), where the unpopulated solution
is stable in white (region 2), the region where populated and unpopulated coexist
is shaded (region 3), and finally the region where the populated solutions is
unstable to patterns in yellow (region 4). Note that the patterns arising
from the MI extend beyond this region and may coexist with the populated or
unpopulated solutions. T stands for the transcritical bifurcation at
$\omega_{d0}/\omega_b=1$, and $SN_1$ for the saddle-node bifurcation
where the subcritical populated solutions ends.}
\label{Phasediagram}
\end{figure}

\newpage
\begin{figure}[H]
  \makeatletter
  \renewcommand{\thefigure}{S\@arabic\c@figure}
  \makeatother
 \includegraphics[width=1\textwidth]{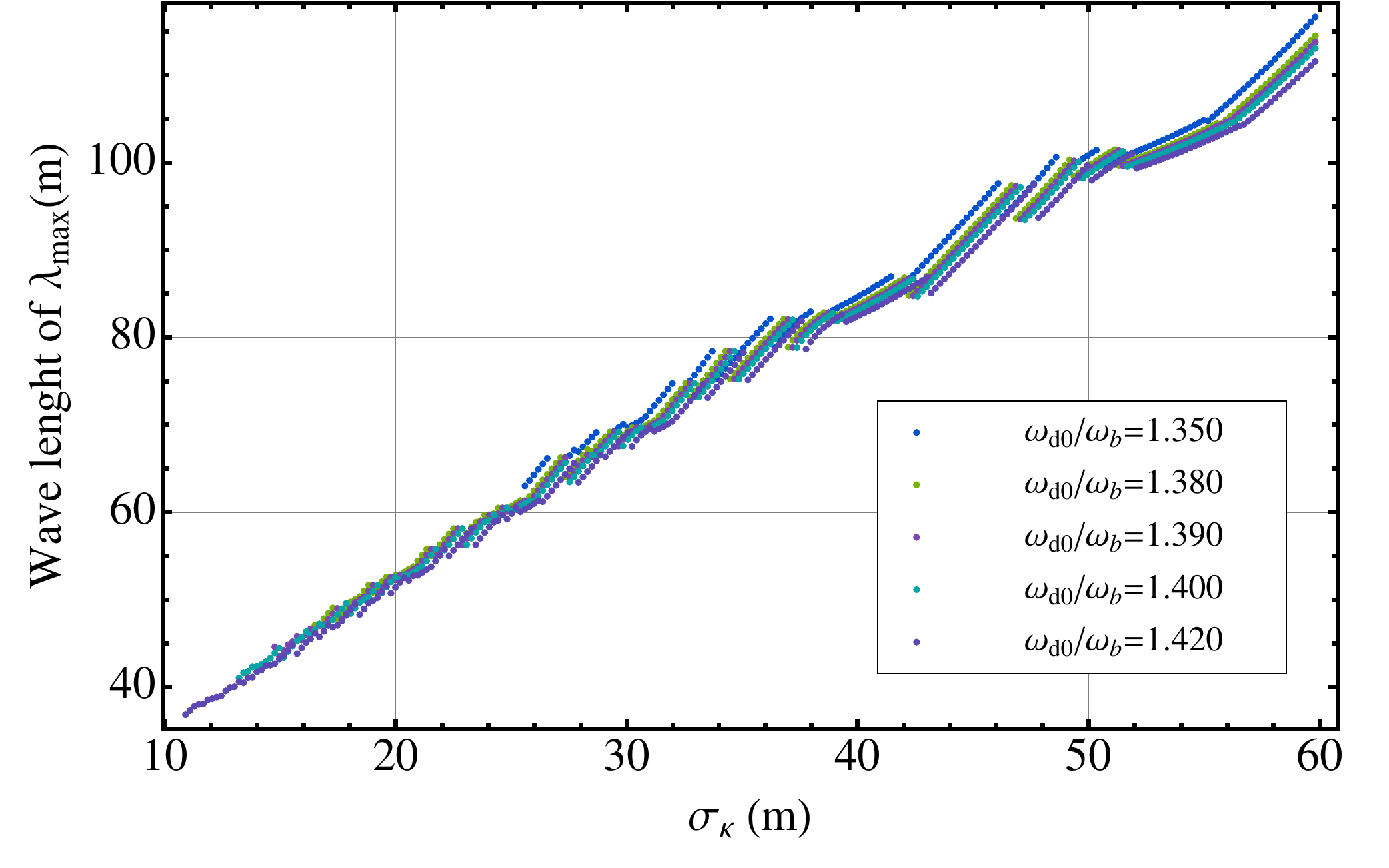}
\caption{Wavelength of the maximum growth rate as function of the competition
range $\sigma_{\kappa}$ for five different values of the intrinsic mortality
$\omega_{d0}$. The parameters are $\omega_b = 0.06$ $ year^{-1}$, $\nu=6.11$
$ cm/year$, $\rho=2.87$ $cm$, $\phi_b=45$, $b=1.25$ $ cm^4year^{-1}$,
$\kappa = 0.048$ $ year^{-1}$, $a=27.38$ $ cm^2$, and $\sigma_\mu=203.7$ $cm$.}
\label{Wavenumdependence}
\end{figure}

\newpage
\begin{figure}[H]
  \makeatletter
  \renewcommand{\thefigure}{S\@arabic\c@figure}
  \makeatother
\includegraphics[width=1\textwidth]{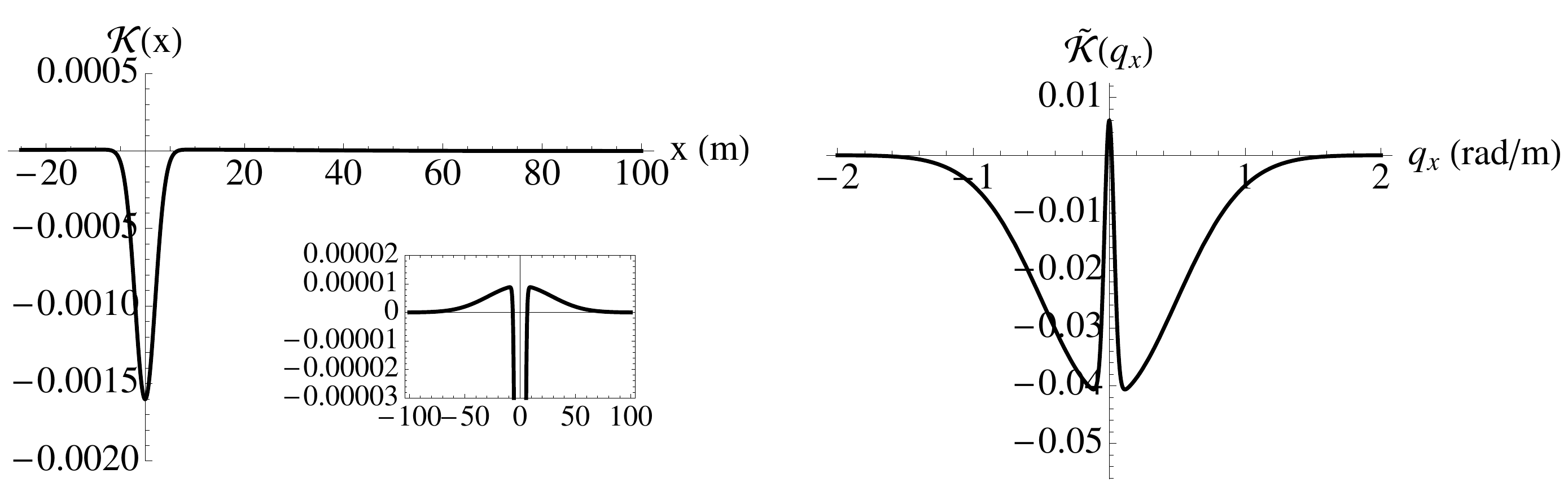}
\caption{Shape of the kernel $\mathcal{K}(\vec{r})$ in real space (left)
and Fourier space (right). A cut in the $x$ and $q_x$ directions for $y=0$
and $q_y=0$ is shown respectively. Here $\kappa=0.048$ $year^{-1}$,
$\omega_{d0}=0.042$ $year^{-1}$, $\sigma_\kappa=2851.4$ $cm$, and $\sigma_\mu= 203.7$ $cm$.}
\label{kernel}
\end{figure}

\newpage
\begin{figure}[H]
  \centering
  \makeatletter
  \renewcommand{\thefigure}{S\@arabic\c@figure}
  \makeatother
 \includegraphics[width=0.85\textwidth]{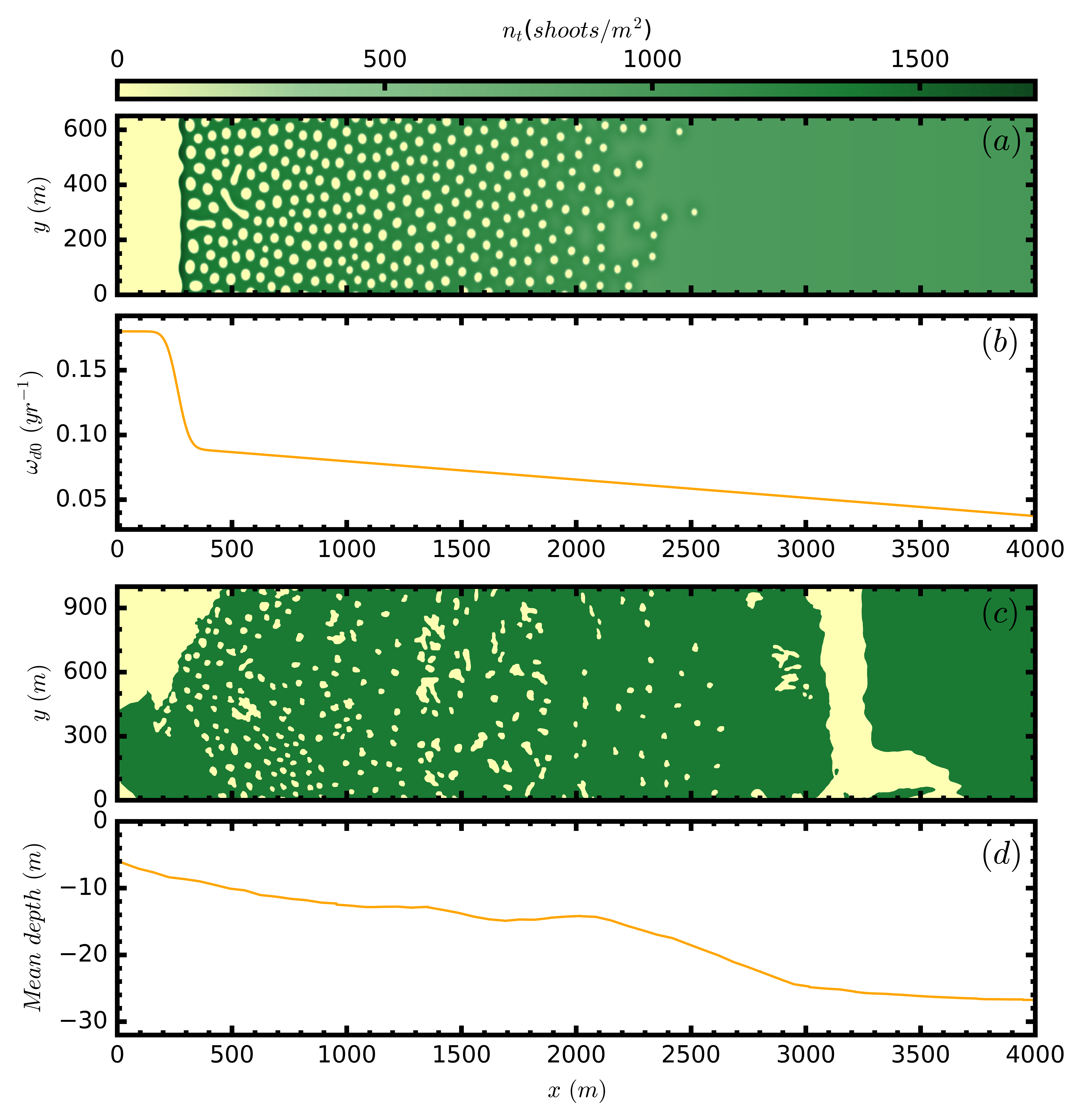}
\caption{Comparison of numerical simulations with patterns in real meadows
in absence of noise in the profile corresponding to the same conditions as
in Fig. 4 (a to d).  Panel (a) shows the final spatial density distribution of
shoots obtained from a numerical simulation of the model in the presence of the
mortality profile plotted in panel (b). (c) Side-scan cartography showing the
presence of \textit{P.oceanica} in the area limited by the following coordinates:
39$^\circ$ 45'54.1"N 3$^\circ$ 09'49.5"E; 39$^\circ$ 47'25.6"N 3$^\circ$ 11'48.7"E;
39$^\circ$ 47'48.6"N 3$^\circ$ 11'19.0"E; 39$^\circ$ 46'17.1"N 3$^\circ$ 09'19.9"E.
Panel d) shows the mean depth of the water in this region averaged over the
$y$ direction. Here irregularities in the patterns are due to the presence of
many competing modes due to the large size of the domain. Due to the absence
of noise the holes are much rounder than in the simulations with noise shown
in the main text.}
\label{meadow}
\end{figure}

\newpage
\begin{figure}[H]
  \makeatletter
  \renewcommand{\thefigure}{S\@arabic\c@figure}
  \makeatother
	\includegraphics[width=1\textwidth]{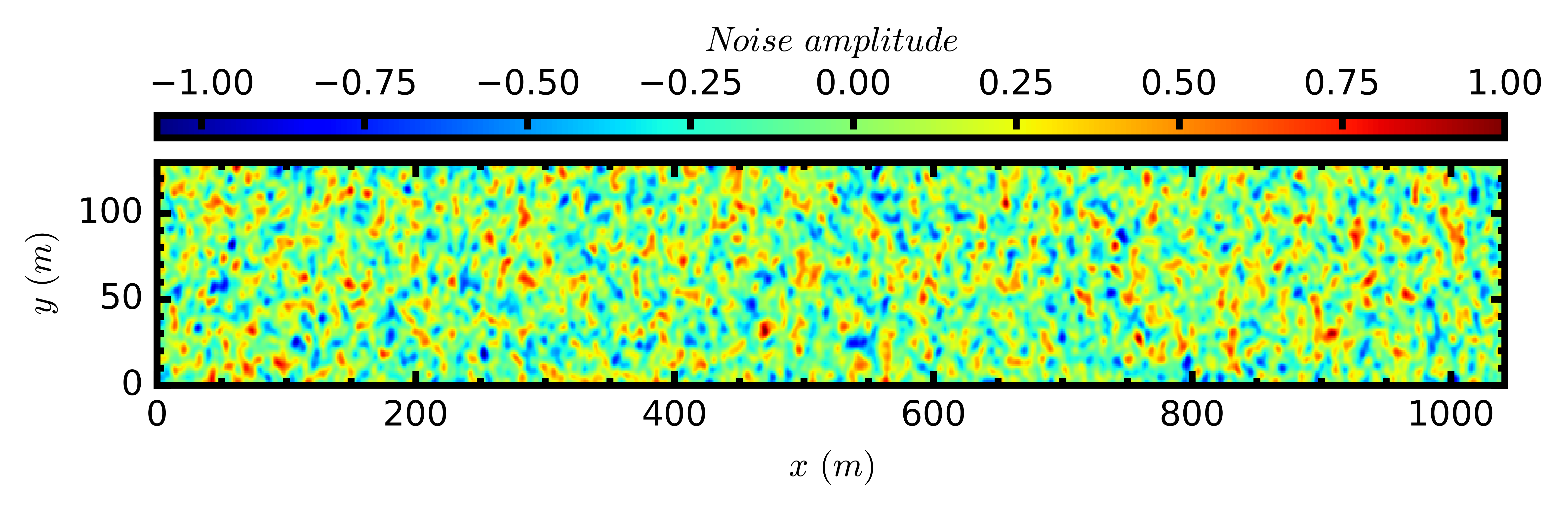}
\caption{Example of the noise distribution used in the numerical simulations.
We take $s=101.83$ $m$ and number of grid points used in each direction is $N_x= 1024$, $N_y=128$.}
\label{NWNoise}
\end{figure}

\newpage
\begin{figure}
  \makeatletter
  \renewcommand{\thefigure}{S\@arabic\c@figure}
  \makeatother
\includegraphics[width=1\textwidth]{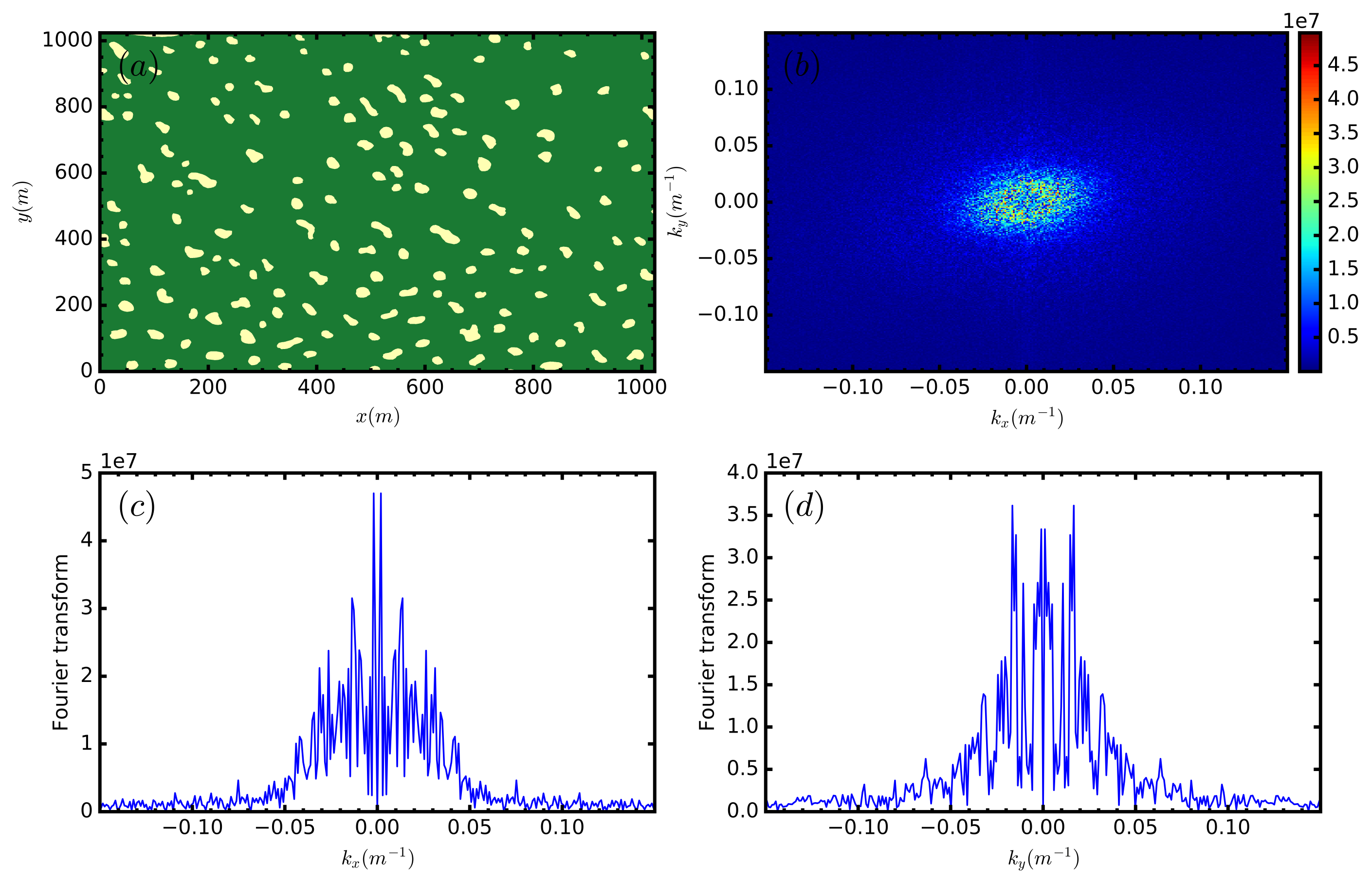}
\caption{Fourier transform of a side-scan cartography image of a rectangular
region of a meadow of \textit{P. oceanica}. Panel (a) shows the presence
(absence) in green (yellow) of Posidonia in a portion of a meadow. Panel (b)
shows the Fourier transform of panel (a). Panels (c)  and (d) are cuts at
$k_y=0$ and $k_x=0$ respectively of panel (b) showing clear peaks at $\left|k\right|\approx 0.017$.}
\label{Fourier}
\end{figure}

\newpage
\begin{figure}
  \centering
  \makeatletter
  \renewcommand{\thefigure}{S\@arabic\c@figure}
  \makeatother
\includegraphics[width=0.85\textwidth]{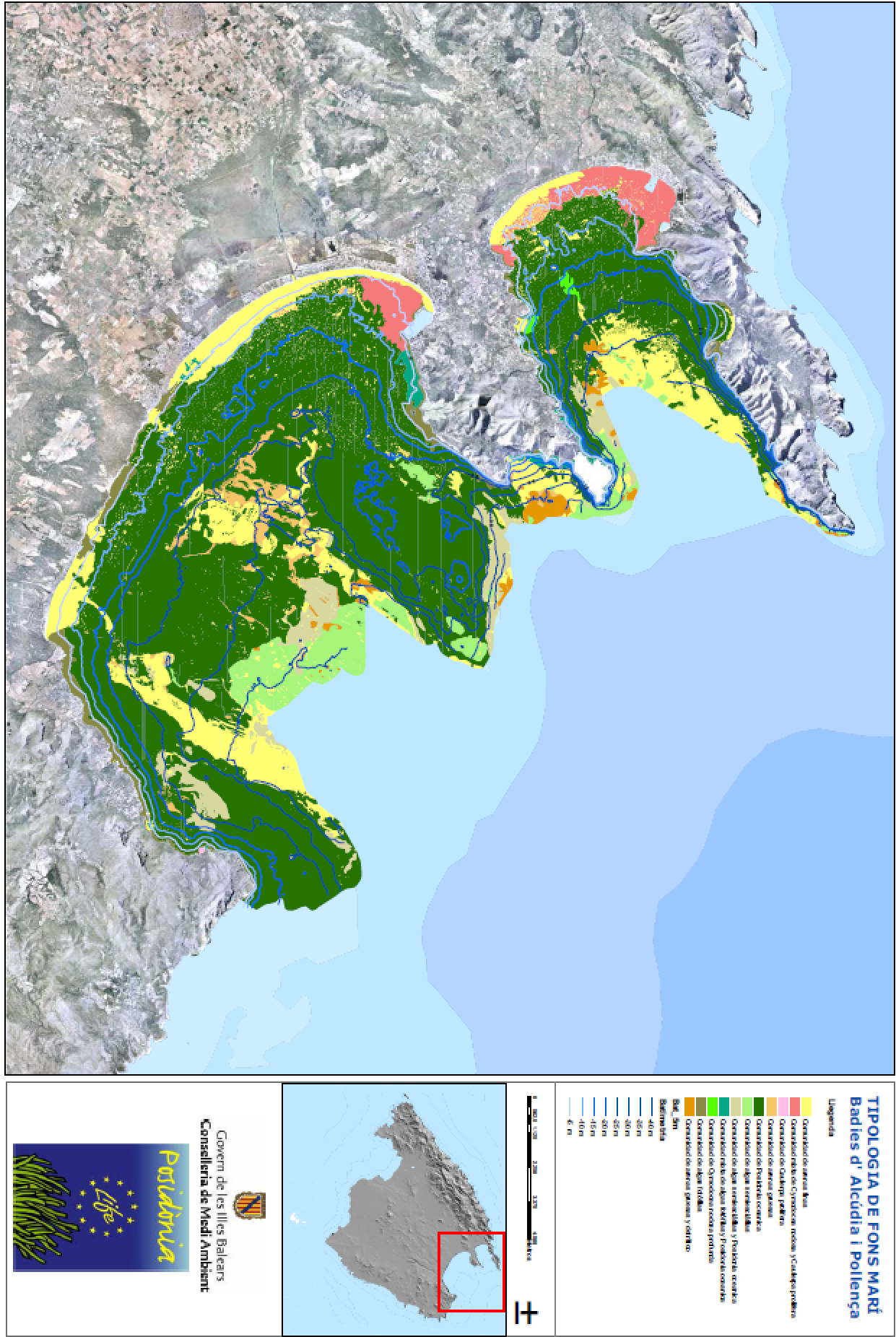}
\caption{Side-scan cartography of Pollen\c{c}a and Alc\'udia bays (Mallorca
Island, Western Mediterranean). This cartography is available in
\href{http://lifeposidonia.caib.es}{http://lifeposidonia.caib.es}.}
\label{Fourier}
\end{figure}

\newpage
\begin{figure}
  \centering
  \makeatletter
  \renewcommand{\thefigure}{S\@arabic\c@figure}
  \makeatother
\includegraphics[width=0.85\textwidth]{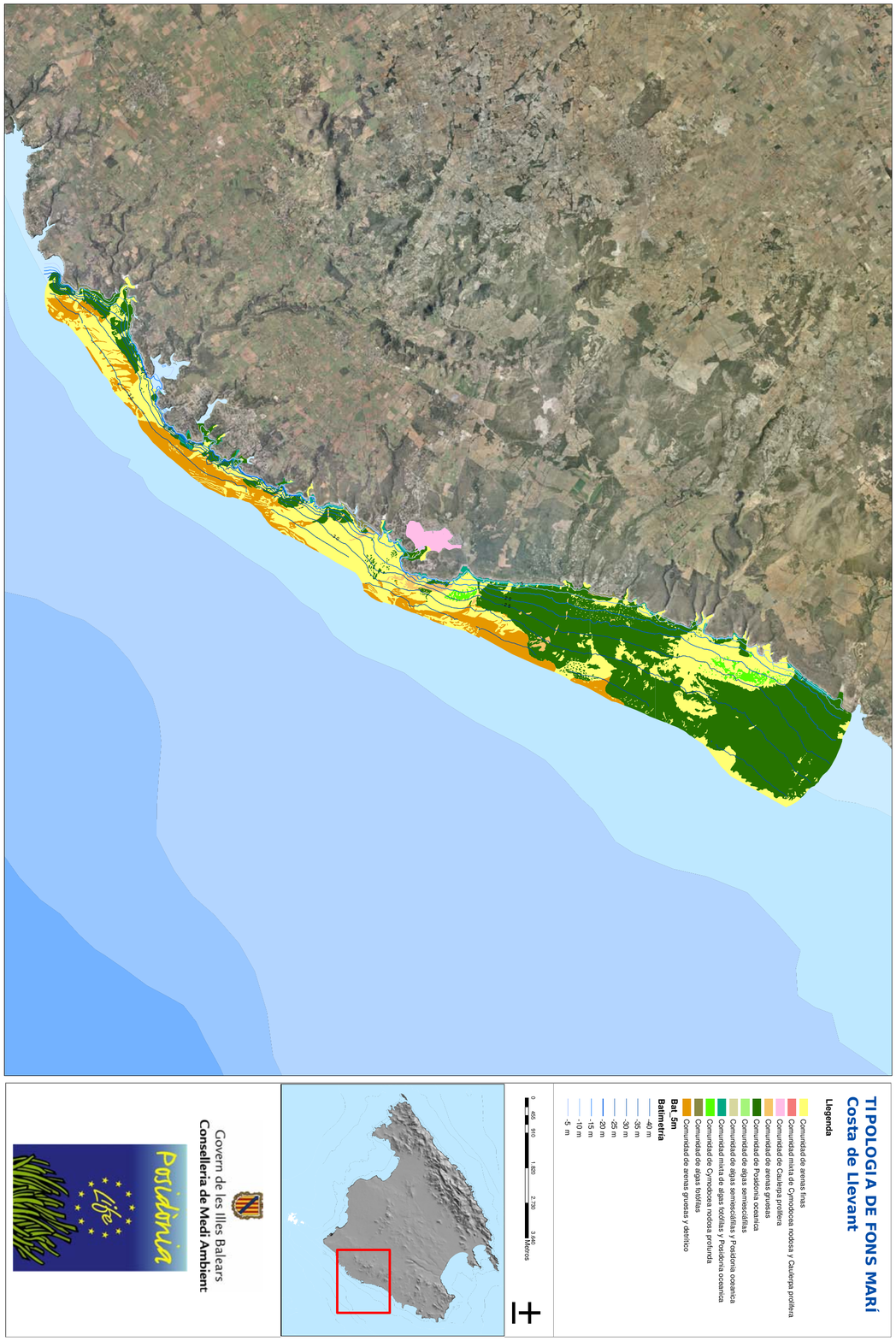}
\caption{Side-scan cartography of Llevant coast (Mallorca Island,
Western Mediterranean). This cartography is available in
\href{http://lifeposidonia.caib.es}{http://lifeposidonia.caib.es}.}
\label{Fourier}
\end{figure}

\newpage
\begin{figure}
  \centering
  \makeatletter
  \renewcommand{\thefigure}{S\@arabic\c@figure}
  \makeatother
\includegraphics[width=0.85\textwidth]{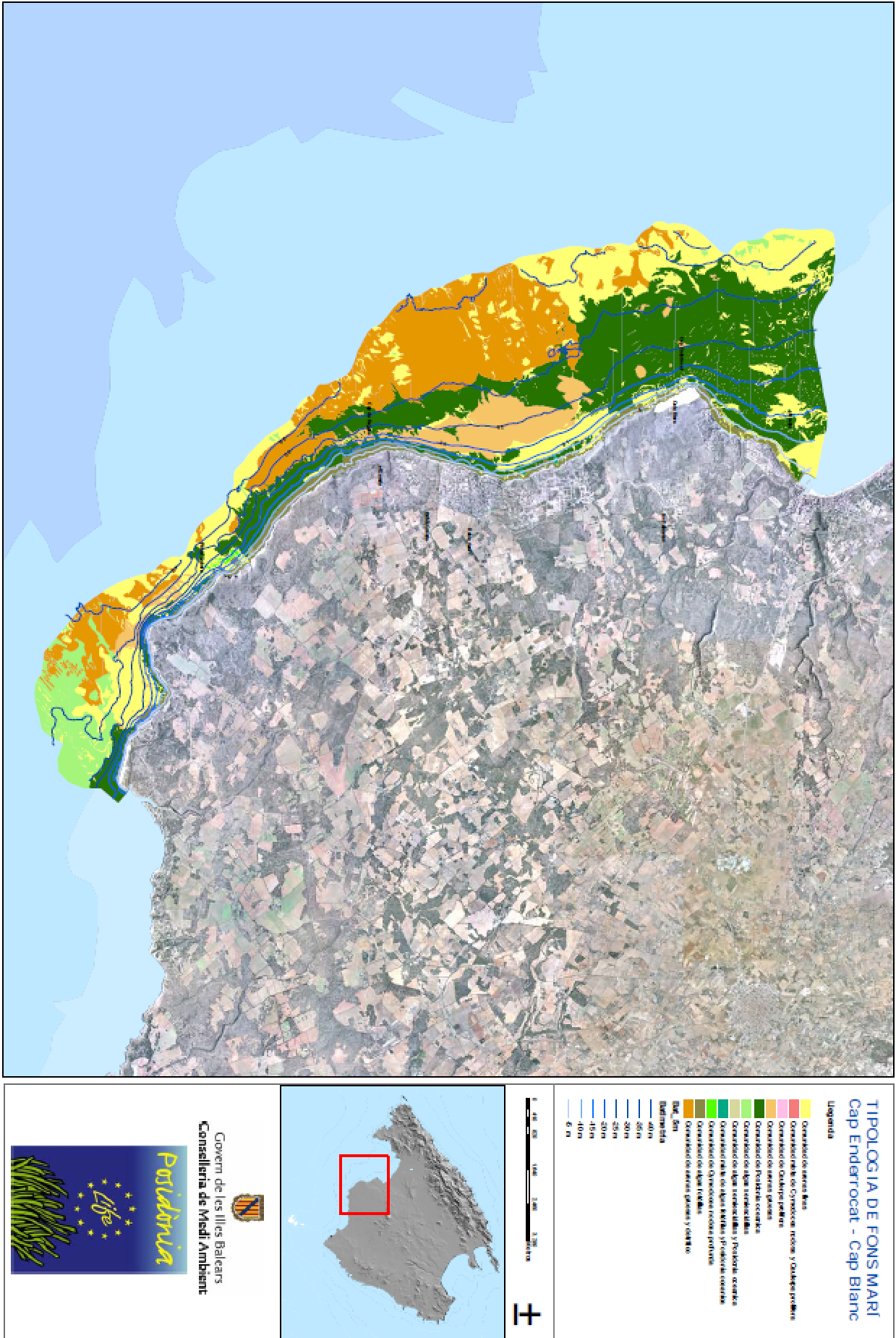}
\caption{Side-scan cartograhpy of Cap enderrocat (Mallorca Island,
Western Mediterranean). This cartography is available in
\href{http://lifeposidonia.caib.es}{http://lifeposidonia.caib.es}.}
\label{Fourier}
\end{figure}

\clearpage
\begin{table}[H]
 \centering
 \caption{Coordinates of analyzed regions. The table
 shows the coordinates of the vertexes of the rectangular
 regions used to measure the characteristic wavelength of patterns in Posidonia meadows.}
\label{table1}
\begin{tabular}{ccccc}
$Lat_0$  & 	$Lon_0$  & 	$Lat_1$  & 	$Lon_1$ \\
\hline
\hline
39$^\circ$52'58.5"N & 3$^\circ$07'43.2"E & 39$^\circ$52'25.2"N &
3$^\circ$08'26.2"E\\
39$^\circ$53'31.8"N & 3$^\circ$05'33.9"E & 39$^\circ$52'58.6"N &
3$^\circ$06'16.9"E \\
39$^\circ$53'31.8"N & 3$^\circ$06'17.0"E & 39$^\circ$52'58.5"N &
3$^\circ$07'00.1"E \\
39$^\circ$49'55.6"N & 3$^\circ$09'52.1"E & 39$^\circ$49'22.4"N &
3$^\circ$10'35.1"E \\
39$^\circ$48'49.2"N & 3$^\circ$10'13.4"E & 39$^\circ$48'15.9"N &
3$^\circ$10'56.4"E \\
39$^\circ$47'09.6"N & 3$^\circ$09'08.6"E & 39$^\circ$46'36.4"N &
3$^\circ$09'51.6"E \\
39$^\circ$47'09.6"N & 3$^\circ$09'51.7"E & 39$^\circ$46'36.3"N &
3$^\circ$10'34.6"E \\
39$^\circ$47'11.0"N & 3$^\circ$10'34.7"E & 39$^\circ$46'36.2"N &
3$^\circ$11'17.7"E \\
39$^\circ$45'46.4"N & 3$^\circ$11'39.1"E & 39$^\circ$45'13.1"N &
3$^\circ$12'22.0"E \\
39$^\circ$45'29.7"N & 3$^\circ$12'22.1"E & 39$^\circ$44'56.4"N &
3$^\circ$13'05.0"E \\
39$^\circ$45'46.3"N & 3$^\circ$12'22.1"E & 39$^\circ$45'13.0"N &
3$^\circ$13'05.0"E \\
\hline
\end{tabular}
\end{table}

\begin{table}[H]
 \centering
 \caption{Measured wavelength of the patterns. The table presents
 observed wavenumber in the rectangular regions indicated in Table \ref{table1}.}
\label{table2}
\begin{tabular}{c}
Wave number ($m^{-1}$)\\
\hline
\hline
 0.017\\
 0.018\\
 0.015\\
 0.016\\
 0.015\\
 0.020\\
 0.019\\
 0.019\\
 0.011\\
 0.01\\
 0.015\\
\hline
\end{tabular}
\end{table}

\newpage
\pagebreak[4]

\paragraph{Movie S1}
Formation of negative hexagons. Movie showing the formation of
a pattern of negative hexagons or fairy circles starting from a
noisy populated solution as initial condition. The parameters
are the same than in Fig. 2 with $\omega_{d0}/\omega_b=1.45$.
The fast evolution of the system at the beginning is slowed
down, as the time indicates, in order to perceive the changes
in the pattern as it forms.

\paragraph{Movie S2}
Formation of stripes. Movie showing the formation of a stripe
pattern starting from a pattern of unstable negative hexagons.
The parameters are the same as in Fig. 2 with
$\omega_{d0}/\omega_b=2.3$. The fast evolution of the system at
the beginning is slowed down, as the time indicates, in order
to perceive the changes in the pattern as it forms.

\paragraph{Movie S3}
Formation of positive hexagons. Movie showing the formation of
positive hexagons starting from a pattern of unstable negative
hexagons. The parameters are the same as in Fig. 2 with
$\omega_{d0}/\omega_b=2.65$. The fast evolution of the system
at the beginning is slowed down, as the time indicates, in
order to perceive the changes in the pattern as it forms.

\paragraph{Movie S4}
Temporal evolution of Fig. 4a. Movie showing the time evolution
of Fig. 4a starting from a noisy homogeneous initial condition.
The evolution of the system, fast at the beginning of the
movie, is slowed down in order to perceive the changes in the
pattern as it forms.

\paragraph{Movie S5}
Temporal evolution of Fig. 4 (e, f, and h). Movie showing the
time evolution of Fig. 4(e, f, and h) starting from a noisy
homogeneous initial condition. The evolution of the system,
fast at the beginning of the movie, is slowed down in order to
perceive the changes in the pattern as it forms.

\paragraph{Movie S6}
Temporal evolution of Fig. 5a. Movie showing the time evolution
of Fig. 5a starting from a noisy homogeneous initial condition.
The evolution of the system, fast at the beginning of the
movie, is slowed down in order to perceive the changes in the
pattern as it forms.

%\bibliographystyle{ScienceAdvances}
%\bibliography{Ref}

\end{document}